# Giant Room-Temperature Third-Order Electrical Transport in a Thin-Film Altermagnet Candidate

Hongyu Chen[1,3], Peixin Qin[1,3]*, Ziang Meng[1,3], Guojian Zhao[1,3], Kai Chen[4], Chuanying Xi[5], Xiaoning Wang[1,3], Li Liu[1,3], Zhiyuan Duan[1,3], Sixu Jiang[1,3], Jingyu Li[1,3], Xiaoyang Tan[1,3], Jinghua Liu[1,3]*, Jianfeng Wang[2]*, Huiying Liu[2]*, Chengbao Jiang[1,3]*, and Zhiqi Liu[1,3]*

[1]School of Materials Science and Engineering, Beihang University; Beijing 100191, China.

[2]School of Physics, Beihang University; Beijing 100191, China.

[3]State Key Laboratory of Tropic Ocean Engineering Materials and Materials Evaluation, Beihang University; Beijing 100191, China

[4]National Synchrotron Radiation Laboratory, University of Science and Technology of China; Hefei, Anhui, 230029, China.

[5]Anhui Key Laboratory of Low-Energy Quantum Materials and Devices, High Magnetic Field Laboratory, Hefei Institutes of Physical Science, Chinese Academy of Sciences; Hefei, Anhui 230031, China

*Corresponding authors. Emails: qinpeixin@buaa.edu.cn; 09077@buaa.edu.cn; wangjf06@buaa.edu.cn; liuhuiying@buaa.edu.cn; jiangcb@buaa.edu.cn; zhiqi@buaa.edu.cn

**Abstract**

Quantum geometry, a quantum mechanical quantity comprised of Berry curvature and quantum metric, describes the geometric structure of the electronic bands in solids. The correlation between nontrivial quantum geometry and quantum materials leads to new findings in condensed matter systems. Here we demonstrate that altermagnets, with spontaneously broken time-reversal ($\mathcal{T}$)–half-lattice-translation and parity-time symmetry, host both $\mathcal{T}$-odd and $\mathcal{T}$-even quantum geometric quantities that simultaneously manifest themselves despite the vanishing net magnetization. Consequently, giant room-temperature third-order electrical transport responses with sizable quantum geometric contributions are observed in (101)-oriented $RuO_2$ thin films, an altermagnetic candidate; in particular, the third-order Hall effect is intimately correlated with altermagnetic order and can serve as a promising tool for detecting the Néel vector. Our work not only supports the existence of altermagnetism in 8-nm-thick $RuO_2$ thin films, but also shows altermagnets as a versatile platform for exploring quantum geometry and constructing quantum electronic and spintronic devices.

**Main text**

Quantum geometry characterizes the distance between eigenstates of electrons in crystalline solids[1–3]. Specifically, quantum metric[4], the real part of the quantum geometric tensor, describes the amplitude difference, while Berry curvature[5], the imaginary part of the quantum geometric tensor, captures the phase variation. According to the parity under time reversal ($\mathcal{T}$), quantum geometry and its multipoles, *i.e.*, spatial distribution of quantum metric and Berry curvature, can be divided into $\mathcal{T}$-even and $\mathcal{T}$-odd categories. The latter one is particularly intriguing owing to its intimate relation with magnetism. In recent years, quantum geometric quantities have been stimulating growing research interest since they not only reveal the geometric structure of Bloch functions, but also underpin various exotic physical phenomena[1–3], including but not limited to the quantum (anomalous) Hall effect[6], flat-band superconductivity[7], and nonlinear optical and transport responses[8,9]. Nevertheless, current experimental explorations on quantum geometric effects have been primarily focusing on two-dimensional materials with either $\mathcal{T}$ or parity-time ($\mathcal{PT}$) symmetry[1–3], which limits the types of spontaneously emergent quantum geometric quantities, thereby hindering the simultaneous manifestation of both $\mathcal{T}$-even and $\mathcal{T}$-odd categories.

In this regard, the recently discovered altermagnetism[10–12] constitutes a novel platform for exploring quantum geometry. The magnetic sublattices of an altermagnet are connected by the combined operation of crystal rotation ($\mathcal{R}$) and $\mathcal{T}$ rather than $\mathcal{T}t_{1/2}$ ($t_{1/2}$ denotes half-unit-cell translation) or $\mathcal{PT}$, so that the sublattice moment arranges in an alternating manner and compensates each other. In correspondence with the $\mathcal{RT}$ symmetry in real space, the electronic band structures of altermagnets exhibit alternating spin polarization and nonaccidental nodal surfaces in reciprocal space in the nonrelativistic limit. Such unique band features serve as natural cradles of diverging facets of quantum geometry, even when spin-orbit coupling is included. Although the anomalous Hall effect (AHE) in altermagnets has been under active investigation[13–17], its emergence can be constrained by additional crystalline symmetries[13,14,17]. Therefore, experimental investigations on other quantum geometric effects rooted in altermagnets, in particular, $\mathcal{T}$-odd transport responses that are capable of characterizing the Néel vector, are of crucial importance in both scientific research and practical applications.

Here we show that thin films of $RuO_2$, a rutile-structured *d*-wave altermagnet candidate that forbids the AHE[13,14], simultaneously exhibit giant third-order longitudinal and transverse electrical transport responses at room temperature. Although the altermagnetism of $RuO_2$ is currently under debate (Supplementary Note 1), the observed anisotropic nonlinear transport behavior, and especially the nonlinear Hall effect with a $\mathcal{T}$-odd feature, are consistent with the presence of long-range altermagnetic order in its epitaxial thin films. Combining symmetry analyses, scaling analyses, and first-principles calculations, we reveal that the nonlinear transport phenomena are related to the net quadrupoles of both quantum metric and Berry curvature, second-order Berry curvature generated by electric fields, and extrinsic disorder scattering mechanisms.

**Symmetry analyses**

We first elaborate on how the symmetry of $RuO_2$ in the altermagnetic state enables nontrivial quantum geometry. As illustrated in Fig. 1a, altermagnetic $RuO_2$ crystallizes in the magnetic point group 4'/*mm'm* with its altermagnetism ensured by $C_{4z}\mathcal{T}$, wherein $C_{4z}$ denotes a four-fold screw rotation with respect to the *c* axis. The intrinsic $\mathcal{T}t_{1/2}$ and $\mathcal{PT}$ symmetry breaking leads to, even in the presence of $\mathcal{P}$ symmetry (Fig. 1a), the spontaneous emergence of $\mathcal{T}$-odd quantum geometric quantities such as Berry curvature and electric-field-induced second-order Berry curvature (2BC) (ref. [18]) in addition to the $\mathcal{T}$-even quantum metric, which distinguishes altermagnetic $RuO_2$ from conventional antiferromagnets[19–25]. Moreover, the quasi-nodal surfaces ($k_{100} = 0$ and $k_{010} = 0$) imposed by $C_{4z}\mathcal{T}$ symmetry (Fig. 1c) serve as natural hotspots of these quantities.

Note that altermagnetism in bulk $RuO_2$ is currently indeterminate, while it is practical to realize altermagnetic order in thin layers of $RuO_2$ (Supplementary Note 1). Therefore, we focus on (101)-oriented epitaxial $RuO_2$ thin films in our study (Fig. 1b). As shown in Extended Data Figs. 1 and 2, thin layers of $RuO_2$ are much easier to exhibit altermagnetism than bulk according to density functional theory calculations. Moreover, an anisotropic exchange-bias effect and temperature-dependent x-ray magnetic linear dichroism (XMLD) are observed in our samples (Extended Data Fig. 3 and Supplementary Note 2). These results clearly support the existence of the envisaged altermagnetism in (101)-$RuO_2$ thin films, and are consistent with the discoveries in recent theoretical and experimental studies[26–30]. Compared to bulk, altermagnetic (101)-RuO$_2$ thin films inherit the broken $\mathcal{PT}$ and $\mathcal{T}t_{1/2}$ symmetry, but possess reduced symmetry elements owing to finite thickness, which only include an inversion center $\mathcal{P}$ (can be absent at surface and interface, Extended Data Fig. 4a), a glide mirror plane $\widetilde{\mathcal{M}}_{010}$, and a two-fold screw rotation axis $\widetilde{C}_{010}^2$ (see Supplementary Note 3 for details). The calculated distribution of quantum metric, Berry curvature, and 2BC in momentum space for a (101)-RuO$_2$ four-layer slab is displayed in Fig. 1d–f, whose hotspots show features of the projected quasi-nodal surfaces in the presence of spin-orbit coupling.

Quantum geometric quantities can manifest themselves in nonlinear electrical transport properties. The preserved $\widetilde{\mathcal{M}}_{010}$ and $\mathcal{P}$ symmetry in (101)-RuO$_2$ thin films prohibit the emergence of the AHE and second-order effects, making third order as the leading-order nonlinear responses (note that $\mathcal{P}$ symmetry breaking at surface/interface can additionally allow $\mathcal{T}$-odd quantum metric dipoles and $\mathcal{T}$-even Berry curvature dipoles, resulting in second-order transport phenomena[20–23,31–35]). In particular, $\mathcal{T}$-odd Berry curvature quadrupoles (BCQs) and 2BC can generate the third-order Hall effect[24,25,36,37] despite the vanishing net magnetization, while $\mathcal{T}$-even quantum metric quadrupoles (QMQs) can give rise to third-order longitudinal and transverse transport[24,37–41]. Therefore, the third-order Hall effect simultaneously incorporates both $\mathcal{T}$-odd and $\mathcal{T}$-even quantum geometric contributions, a hallmark of altermagnets. After flipping the Néel vector of RuO$_2$, third-order longitudinal transport would remain invariant (Fig. 1g), while the third-order Hall effect induced by $\mathcal{T}$-odd mechanisms would reverse sign (Fig. 1h,i). Moreover, the orientation of Néel order dictates the crystalline anisotropy of the third-order Hall effect (Supplementary Note 3), making it a promising electric probe to altermagnetic structures.

**Third-order electrical transport in (101)-RuO$_2$ thin films**

We next study the quantum geometry through nonlinear transport measurements. The basic structural characterization on the (101)-RuO$_2$ thin films grown on (101)-TiO$_2$ substrates can be found in our previous work[42]. We pattern 8-nm-thick plain thin films into Hall bars to study the electrical transport properties. The optical image of a typical device is displayed in the inset of Fig. 2c. Figure S1a depicts the temperature ($T$)-dependent dc resistivity ($\rho_{xx}$) at 50–400 K of a typical device. The RuO$_2$ layer exhibits a metallic transport behavior with d$\rho_{xx}$/d$T$ > 0 and a room-temperature $\rho_{xx}$ of ~65.6 μΩ cm, comparable to those in other studies[29,43]. As shown in Extended Data Fig. 4, we first investigate the second-order transport of the samples. The observed second-order Hall effect is confirmed to be of interfacial origin according to its pronounced thickness dependence. A more detailed discussion can be found in Supplementary Note 4.

The third-order transport properties are investigated by measuring the amplitude of the third-harmonic longitudinal (transverse) voltages $V_x^{3\omega}$ ($V_y^{3\omega}$) generated by an applied ac current of $I_x^\omega \sin(\omega t)$, wherein $I_x^\omega$ and $\omega$ are the amplitude and angular frequency, respectively (Fig. 2a,e). As shown in Fig. 2b,c, $V_x^{3\omega}$ scales with the cube of $I_x^\omega$ in third-order longitudinal transport. In addition, the recorded $V_x^{3\omega}$ signals are close in magnitude for $I_x^\omega$ applied along [010], [$\bar{1}$01] and [$\bar{1}$11]$^*$ (a direction slightly tilted away from [$\bar{1}$11] of the tetragonal crystal structure). We confirm that the nonzero $V_x^{3\omega}$ is irrelevant to extrinsic artifacts such as the thermoelectric effect, contact junctions, and capacitive coupling, since the $V_x^{3\omega}$ ($I_x^\omega$) relation exhibits negligible dependence on the measurement protocols (Fig. 2b,d). A more detailed discussion can be found in Supplementary Note 5. It is worth noting that the $V_x^{3\omega}$ detected with different electrode pairs in the same Hall bar are almost identical (Fig. 2c), despite their possibly distinct domain population and (or) distribution. This indicates the $\mathcal{T}$-even feature of the third-order longitudinal transport.

We next move on to the third-order Hall effect. As displayed in Fig. 2f, the $V_y^{3\omega}$ is proportional to the cube of $I_x^\omega$, in line with the third-order nature (see Supplementary Note 7 for the determination of $V_y^{3\omega}$). Notably, the third-order Hall effect exhibits prominent crystalline anisotropy—the generated $V_y^{3\omega}$ is much larger for $I_x^\omega$ applied along [$\bar{1}$11]$^*$ than along [010] or [$\bar{1}$01]. This, together with the additional measurement results in Fig. S5, excludes extrinsic artifacts as the origin of the cubic $V_y^{3\omega}$($I_x^\omega$) relation. In contrast to the third-order longitudinal transport, the third-order Hall responses measured with different Hall crossings in the same device are distinct in magnitude (Fig. 2g). This should be an intrinsic property of the sample as the Hall bar defined by standard lithography is strictly centrosymmetric, and no contact junction is formed. We thus tend to perceive this anomaly as the signature of a $\mathcal{T}$-odd effect—the domain population and (or) distribution in the two crossings are different, so that the $\mathcal{T}$-odd contributions to $V_y^{3\omega}$ are disparate. We note that the second-order transverse transport of the sample also exhibits similar characteristics (Extended Data Fig. 4d), in sharp contrast to the reported case of $\mathcal{T}$-even materials[32].

We further confirm the $\mathcal{T}$-odd feature with field-cooling experiments. Note that the full switching of the Néel vector in RuO$_2$ thin films typically entails a magnetic field larger than 50 T (ref. [44]). Instead, recent studies have indicated that the altermagnetic order of RuO$_2$ thin films can be affected by field cooling from elevated temperature[45–47]. Following these works, we study the third-order transport responses in another device before and after cooling from 423 K (close to or above its Néel point, Supplementary Note 2, but too low to affect the crystallinity of an oxide material) to room temperature in an out-of-plane magnetic field of ~30 T. We emphasize that, owing to compensated magnetization, a static magnetic field in the cooling process could hardly induce coherent reversal of the Néel vector to form a single-domain state, but could rather trigger the repopulation and/or redistribution of altermagnetic domains. In addition, we have confirmed that simple zero-field cooling cannot lead to an observable change in the third-order transport. As displayed in Fig. 2h, the sign of the third-order Hall effect reverses after field cooling. This indicates that the $\mathcal{T}$-odd mechanisms play an important role in engendering the nonlinear responses, and that the population of opposite altermagnetic domains increases, at least in the Hall crossing regions in the device. In contrast, the third-order longitudinal transport is almost invariant after field cooling (Fig. S4c,d), consistent with its prevailing $\mathcal{T}$-even mechanisms.

**Crystalline anisotropy of third-order electrical transport**

We next investigate the crystalline anisotropy of the third-order transport in a more detailed manner. To this end, a twelve-terminal device of symmetric shape is employed (Fig. 3a). As displayed in Fig. 3b–d, we apply $I_x^\omega$ at various angles $\Psi$ relative to [010] and measure the resultant first-harmonic longitudinal voltage $V_x^\omega$ and third-harmonic voltages $V_x^{3\omega}$ and $V_y^{3\omega}$. Figure 3e depicts the $\Psi$ dependence of the first-order longitudinal resistance $R_{//}^\omega$ and third-order nonlinearity $E_x^{3\omega}/(E_x^\omega)^3$ and $E_y^{3\omega}/(E_x^\omega)^3$ extracted from Fig. 3b–d (see Supplementary Note 6 for details), all of which exhibit two-fold symmetry due to the constraints imposed by $\widetilde{\mathcal{M}}_{010}$. Moreover, taking into account the crystalline symmetry of altermagnetic (101)-RuO$_2$, the $\Psi$ dependence can be well fitted by (see Supplementary Note 8 for the derivation)

$$R_{//}^{\omega} = R_b\cos^2\Psi + R_a\sin^2\Psi \tag{1}$$

$$\frac{E_x^{3\omega}}{(E_x^{\omega})^3} = \frac{\sigma_1^{(3)}\rho_b^4\cos^4\Psi + \sigma_2^{(3)}\rho_a^4\sin^4\Psi + 3[\sigma_3^{(3)} + \sigma_4^{(3)}]\rho_b^2\rho_a^2\sin^2\Psi\cos^2\Psi}{(\rho_b\cos^2\Psi + \rho_a\sin^2\Psi)^3} \tag{2}$$

$$\frac{E_y^{3\omega}}{(E_x^{\omega})^3} = \frac{[3\sigma_4^{(3)}\rho_b^2\rho_a^2 - \sigma_1^{(3)}\rho_b^4]\sin\Psi\cos^3\Psi + [\sigma_2^{(3)}\rho_a^4 - 3\sigma_3^{(3)}\rho_b^2\rho_a^2]\sin^3\Psi\cos\Psi}{(\rho_b\cos^2\Psi + \rho_a\sin^2\Psi)^3} \tag{3}$$

wherein $R_b$ ($\rho_b$) and $R_a$ ($\rho_a$) are the resistances (resistivities) of RuO$_2$ along [010] and [$\bar{1}$01], respectively; $\sigma_1^{(3)}$–$\sigma_4^{(3)}$ are the related components of the third-order conductivity tensor. Notably, the third-order Hall effect vanishes when $I_x^{\omega}$ is applied parallel or perpendicular to $\widetilde{\mathcal{M}}_{010}$, which accounts for the prominent anisotropy shown in Fig. 2f. In contrast, third-order longitudinal transport is permitted in all the $\Psi$ and is thus loosely restricted by $\widetilde{\mathcal{M}}_{010}$. Consequently, the anisotropy of the nonlinear transport properties is in good accordance with the symmetry of altermagnetic (101)-RuO$_2$.

**Microscopic mechanisms of third-order electrical transport**

We finally try to shed light on the microscopic mechanisms of the third-order transport with scaling analyses and first-principles calculations. As displayed in Fig. S7a, the Hall carrier density of a typical sample is less dependent on temperature at 200–300 K. We thus employ the nonlinear transport data collected in this temperature range to study the relation between third-order longitudinal and Hall conductivity $\sigma_{//}^{(3)}$ ($I_x^{\omega}$ applied along [010]) and $\sigma_{\perp}^{(3)}$ ($I_x^{\omega}$ applied along [$\bar{1}$11]$^*$) and first-order longitudinal conductivity $\sigma_{xx}$ (see Supplementary Note 6 for the calculation method), so that we can make the approximation of $\tau \propto \sigma_{xx}$ ($\tau$ denotes electron relaxation time).

Since the third-order longitudinal transport exhibits a $\mathcal{T}$-even feature, we fit the $\sigma_{//}^{(3)}(\sigma_{xx})$ curve only taking into account $\mathcal{T}$-even mechanisms. As displayed in Fig. 4a, the relation between $\sigma_{//}^{(3)}$ and $\sigma_{xx}$ can be well captured by

$$\sigma_{//}^{(3)} = \xi\sigma_{xx}^3 + \eta\sigma_{xx} \tag{4}$$

wherein $\xi$ and $\eta$ are fitting parameters. According to the fitting results, the $\xi\sigma_{xx}^3$ and $\eta\sigma_{xx}$ terms are comparable (~6.5 and ~8.1 μm A V$^{-3}$, respectively) at room temperature. The $\xi\sigma_{xx}^3$ term can arise from the Drude-like mechanism and disorder scattering, while the $\eta\sigma_{xx}$ term can be attributed to QMQs and also possible scattering. To investigate the role of quantum geometric quantities, we perform first-principles calculations for a (101)-oriented four-layer RuO$_2$ slab. The calculated results for the QMQ and Drude-like contributions to $\sigma_{//}^{(3)}$ are shown in Fig. 4b. Taking an estimated $\tau$ of ~0.04 ps at room temperature according to the first-order conductivity, the $\sigma_{//,\text{QMQ}}^{(3)}$ induced by QMQ is on the order of 1 μm A V$^{-3}$, while the Drude-like $\sigma_{//,\text{Drude}}^{(3)}$ is two orders of magnitude smaller. Therefore, the $\sigma_{//,\text{QMQ}}^{(3)}$ related to $\eta\sigma_{xx}$ is dominant at room temperature, qualitatively in accordance with the experimental scaling analyses. The quantitative discrepancy may stem from two aspects, *i.e.*, the size effect of the four-layer slab that could limit the magnitude of $\sigma_{//}^{(3)}$, and the contributions from scattering effects, especially skew scattering mechanisms proportional to high orders of $\tau$. Experimentally disentangling these extrinsic mechanisms is challenging owing to the strong variation in carrier density within the full temperature range (Fig. S7a), which deserves further investigation.

For the third-order Hall effect, as shown in Fig. 4c, the relation between $\sigma_\perp^{(3)}$ and $\sigma_{xx}$ can be well described by

$$\sigma_\perp^{(3)} = \zeta\sigma_{xx} + \lambda \qquad (5)$$

without resort to other higher-order $\tau$-dependent components, wherein $\zeta$ and $\lambda$ are fitting constants. According to the fitting results, the $\zeta\sigma_{xx}$ and $\lambda$ terms are comparable (~−4.2 and ~2.7 μm A V$^{-3}$, respectively) at room temperature. Note that a $\zeta\sigma_{xx}$ slightly larger than $\lambda$ indicates that, owing to the multidomain state at the Hall crossing, $\mathcal{T}$-even mechanisms prevail the third-order Hall responses in this specific sample. Meanwhile, our theoretical calculations reveal that the third-order Hall conductivity $\sigma_\perp^{(3)}$ induced by 2BC ($\sigma_{\perp,\text{2BC}}^{(3)}$, contributes to the $\lambda$ term) and QMQs ($\sigma_{\perp,\text{QMQ}}^{(3)}$, contributes to the $\zeta\sigma_{xx}$ term) are dominant quantum geometric contributions at room temperature (Fig. 4d and Fig. S9a). Notably, taking a $\tau$ of ~0.04 ps, $\sigma_{\perp,\text{2BC}}^{(3)}$ and $\sigma_{\perp,\text{QMQ}}^{(3)}$ are comparable (both on the order of 0.1 μm A V$^{-3}$), consistent with the scaling analyses, while the quantitative discrepancy is acceptable considering the size effect. In addition, we also perform scaling analyses with higher-order $\mathcal{T}$-odd mechanisms unincorporated in Eq. 5. The result shows that $\mathcal{T}$-odd contributions proportional to $\tau^4$ and $\tau^2$ are relatively small compared to $\zeta\sigma_{xx}$ and $\lambda$ (Fig. S9b). The former is induced by the second-order skew scattering, and the latter is collectively contributed by BCQs (calculated on the order of 0.01 μm A V$^{-3}$, Fig. S9a) and $\mathcal{T}$-odd scattering effects. We discuss possible scattering contributions to third-order transport, and propose $\mathcal{T}$-odd scattering mechanisms such as second-order skew scattering and side-jump effect in Supplementary Note 9.

Therefore, we conclude that quantum geometry, which is largely contributed by the altermagnetic band structures, plays an important role in giving rise to the observed nonlinear transport responses. In addition, extrinsic scattering mechanisms should also not be disregarded, especially in view of the $\mathcal{T}$-symmetry breaking spin textures in altermagnets that could give rise to additional scattering channels. Finally, although the exchange bias, XMLD, and sign reversal of the third-order Hall effect consistently support the existence of altermagnetic order in our samples, we do not intend to conclusively claim the altermagnetism. In fact, it remains an open question how and within what thickness range altermagnetism can be stabilized in RuO$_2$ thin films. Moreover, even in the nonmagnetic state, RuO$_2$ is a topologically nontrivial metallic oxide with the presence of Dirac nodal lines[48]. It is thus interesting to study whether and to what extent paramagnetic RuO$_2$ can exhibit nonlinear transport effects. In this regard, our further investigation suggests the degradation of altermagnetism with increasing thickness, leading to an attenuated $\mathcal{T}$-odd component of the $\sigma^{(3)}_\perp$ in an 80-nm-thick sample (Supplementary Note 10).

**Conclusions**

In summary, we report the observation of third-order electrical transport properties in epitaxial (101)-RuO$_2$ layers that can be well-interpreted based on altermagnetism. As an altermagnetic candidate with spontaneously broken $\mathcal{PT}$ or $\mathcal{T}t_{1/2}$ symmetry, RuO$_2$ thin films host both $\mathcal{T}$-odd and $\mathcal{T}$-even quantum geometric quantities including BCQs, 2BC, and QMQs, all of which make considerable contributions to the studied nonlinear transport responses. In particular, the third-order transverse transport is found to reverse sign after field cooling due to the repopulation and/or redistribution of altermagnetic domains, which indicates its important $\mathcal{T}$-odd mechanisms. This, together with the prominent crystalline anisotropy, makes the third-order Hall effect a promising probe to altermagnetic order. Looking ahead, we emphasize that the room-temperature third-order transport in RuO$_2$ thin films is giant in magnitude compared with that in other (two-dimensional) topological materials [24,40,41,49–53] (Extended Data Fig. 5). This demonstrates the tremendous potential of RuO$_2$ as a candidate material for emerging quantum electronic and spintronic devices. Moreover, in addition to altermagnets with *d*- or *g*-wave characteristics, we note that diverging aspects of nontrivial quantum geometry are, in principle, allowed in various other unconventional antiferromagnets[54], which is expected to result in exotic quantum geometric physics beyond nonlinear transport.

**Acknowledgements**

The authors sincerely thank Prof. Shengyuan A. Yang at Hong Kong Polytechnic University for fruitful and enlightening discussions on the theoretical understanding of our observation. The authors acknowledge support from the National Natural Science Foundation of China (no. 52401300 to P.Q., nos. 52425106 and 52271235 to Z.L., no. 52121001 to Z.L. and C.J., no. 12304053 to H.L., no. 12474217 to J.W., no. 524B2003 to Z.M., no. 525B2008 to L.L. and no. U25A20244), the Beijing Natural Science Foundation (no. JQ23005 to Z.L. and no. 1242023 to J.W.), the National Key R&D Program of China (nos. 2022YFA1602700 and 2022YFB3506000 to Z.L. and no. 2022YFA1402604 to H.L.) and the Open Research Fund of the Pulsed High Magnetic Field Facility, Huazhong University of Science and Technology (no. WHMFC2024020 to P.Q.). This work is also supported by the Fundamental Research Funds for the Central Universities. The authors acknowledge the Analysis & Testing Center of Beihang University for the assistance. The authors acknowledge the staff members of the XMCD beamline at the NSRL in Hefei (https://cstr.cn/31131.02.HLS) for providing the technical support and assistance in XAS data collection and analysis. The authors acknowledge the WM5 of the Steady High Magnetic Field Facility, CAS (https://cstr.cn/31125.02.SHMFF.WM5) for the assistance on the experiment.

**Author contributions**

H.C., P.Q., C.J., and Z.L. conceived the project. H.C., P.Q., Z.M., G.Z., X.W., L.L., Z.D., S.J, J.Li., and X.T. fabricated the samples with support from J.Liu. H.C., P.Q., Z.M., G.Z., X.W., L.L., Z.D., S.J, J.Li., and X.T. processed the devices and performed electrical transport and magnetic measurements. H.L. and J.W. performed theoretical calculations. K.C. carried out x-ray absorption spectroscopy measurements. C.X. performed field cooling of the sample with a magnetic field of 30 T. H.C., H.L., and J.W. wrote the manuscript with support from P.Q., C.J., and Z.L. Z.L. led and coordinated the project. All authors contributed to the discussion on the results.

**Competing interests**

The authors declare no competing interests.

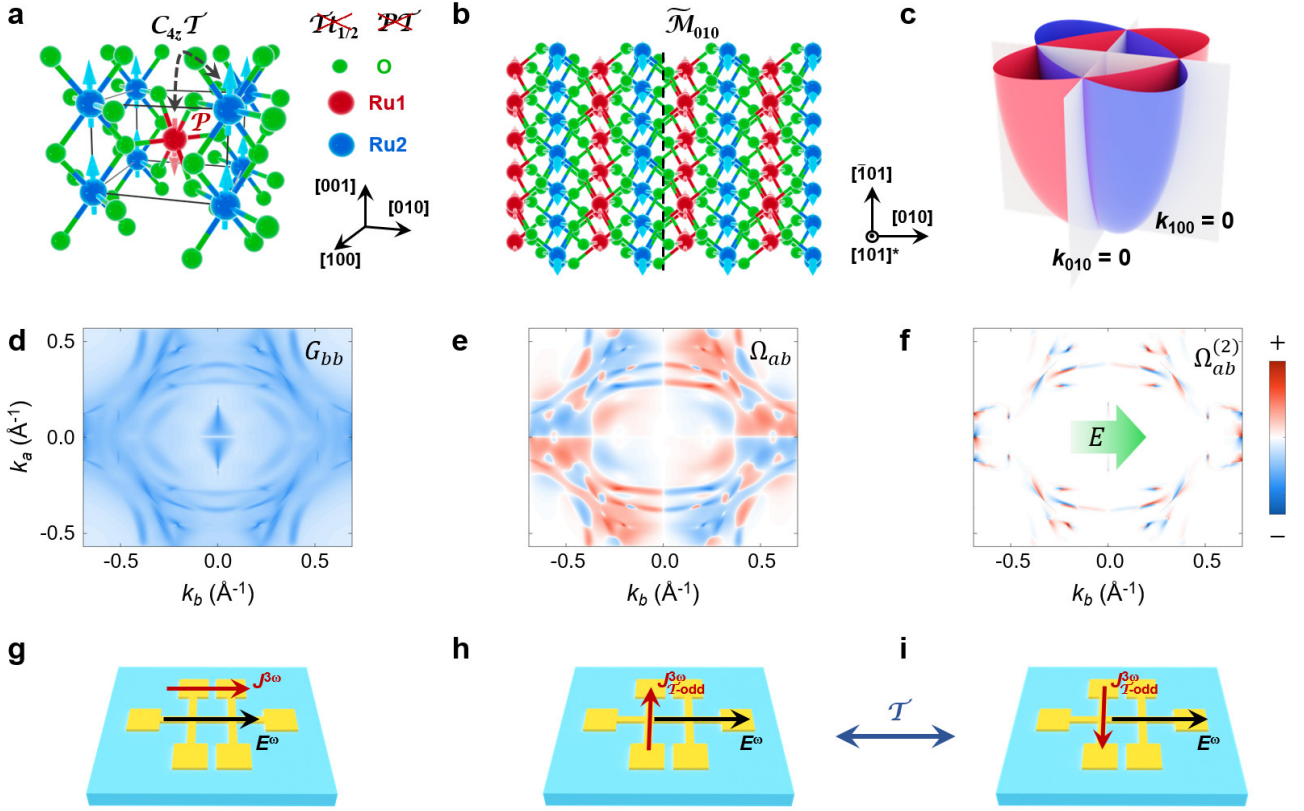

**Fig. 1 | Quantum geometry and third-order electrical transport allowed by altermagnetic symmetry in (101)-oriented RuO₂ thin films. a,** Magnetic crystal structure of altermagnetic RuO₂, which preserves parity ($\mathcal{P}$) symmetry and combined $C_{4z}\mathcal{T}$ symmetry of four-fold screw rotation ($C_{4z}$) and time reversal ($\mathcal{T}$) but breaks $\mathcal{T}t_{1/2}$ ($t_{1/2}$ denotes half-unit-cell translation) and $\mathcal{PT}$ symmetry. The arrows denote the orientation of the magnetic moment. **b,** Top view of a (101)-oriented RuO₂ thin film, which possesses a glide mirror plane $\widetilde{\mathcal{M}}_{010}$. Due to the tetragonal structure of RuO₂, [101]* is slightly tilted away from the [101] direction. **c,** Schematic illustration of the quasi-nodal surfaces enforced by the $C_{4z}\mathcal{T}$ symmetry in RuO₂. **d–f,** Momentum space distribution of quantum metric component $G_{bb}$ (**d**), Berry curvature component $\Omega_{ab}$ (**e**), and second-order Berry curvature $\Omega_{ab}^{(2)}$ induced by an electric field $E$ (**f**) in a four-layer (101)-RuO₂ slab. $b$ and $a$ axes are along [010] and [$\bar{1}$01] of bulk RuO₂, respectively. **g–i,** Schematic illustration of third-order longitudinal (**g**) and transverse (**h** and **i**) transport wherein a third-harmonic current density $J^{3\omega}$ is generated by the applied ac electric filed $E^\omega$. The third-order Hall effect induced by $\mathcal{T}$-odd mechanisms reverses sign after a $\mathcal{T}$ operation.

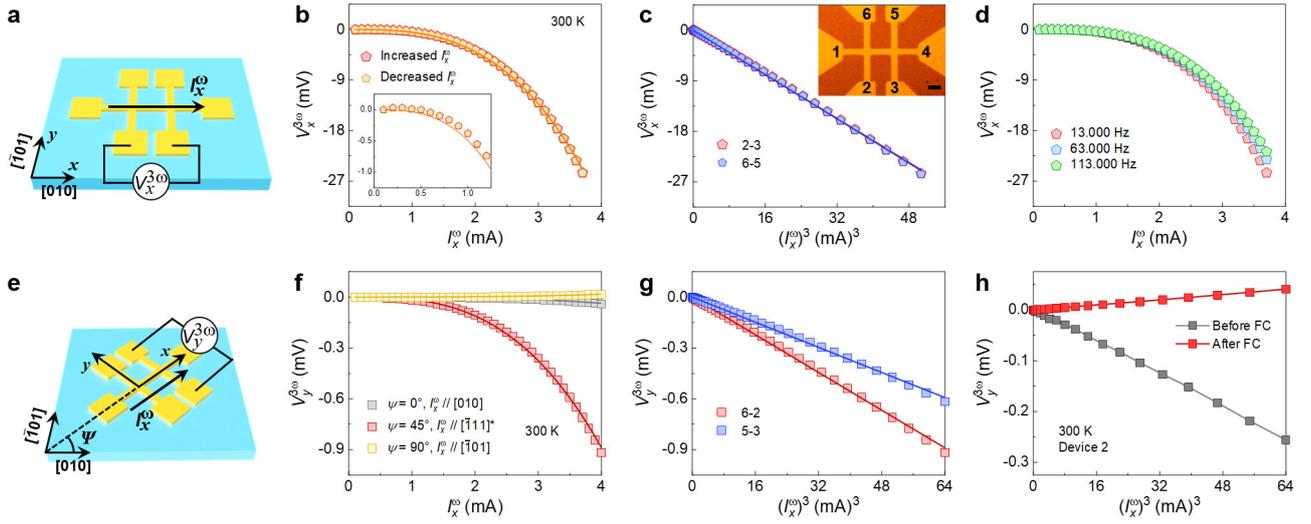

**Fig. 2 | Third-order electrical transport in a (101)-oriented RuO₂ thin film. a**, Schematic illustration on the measurement protocol of third-order longitudinal transport. **b**–**d**, Relation between third-harmonic longitudinal voltages $V_x^{3\omega}$ and the applied ac current $I_x^{\omega}$ measured with increased/decreased current (**b**), different electrode pairs (**c**) and frequency (**d**) at room temperature. $I_x^{\omega}$ is directed along [010]. The inset in **b** displays the data collected with a small $I_x^{\omega}$. The inset in **c** displays the optical image of a typical Hall bar device. The scale bar represents a length of 5 μm. The electrodes are numbered to indicate which pair is employed in the measurement. **e**, Schematic illustration on the measurement protocol of the third-order Hall effect. $\Psi$ denotes the angle between the $x$ axis of the coordinate system and the [010] direction. **f**, Relation between third-harmonic transverse voltages $V_y^{3\omega}$ and $I_x^{\omega}$ at room temperature. $I_x^{\omega}$ is applied at angles $\Psi$ of 0°, 45°, and 90°. **g**, $V_y^{3\omega}$ as a function of $(I_x^{\omega})^3$ at room temperature measured with distinct Hall crossings in the same $[\bar{1}11]^*$-oriented Hall-bar device. $[\bar{1}11]^*$ is slightly tilted away from $[\bar{1}11]$ of the tetragonal structure. **h**, Room-temperature third-order Hall effect in another device measured before and after field cooling (FC) from 423 K. $I_x^{\omega}$ is applied along $[\bar{1}11]^*$. The solid lines in **b** and **f** are cubic fittings to the data. The solid lines in **c** and **g** are linear fittings to the data.

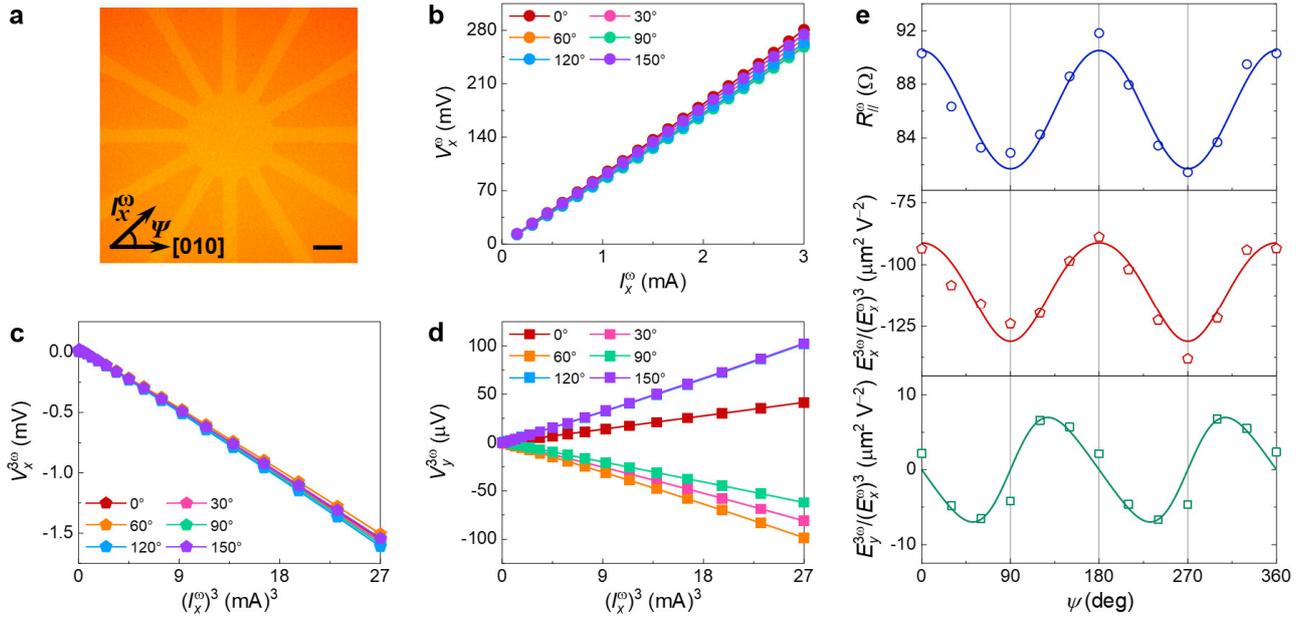

**Fig. 3 | Crystalline anisotropy of the third-order electrical transport in a (101)-oriented RuO$_2$ thin film. a**, Optical image of the twelve-terminal device. The scale bar represents a length of 10 μm. **b–d**, First-harmonic longitudinal voltage $V_x^\omega$ (**b**) and third-harmonic voltage $V_x^{3\omega}$ (**c**) and $V_y^{3\omega}$ (**d**) as a function of $I_x^\omega$ or $(I_x^\omega)^3$ at room temperature. $I_x^\omega$ is applied along different crystalline directions. **e**, $\Psi$ dependence of first-order longitudinal resistance $R_{//}^\omega$, third-order longitudinal nonlinearity $E_x^{3\omega}/(E_x^\omega)^3$, and third-order transverse nonlinearity $E_y^{3\omega}/(E_x^\omega)^3$. The data are extracted from **b–d**. The solid lines in **e** are fittings to the data according to Eqs. 1–3.

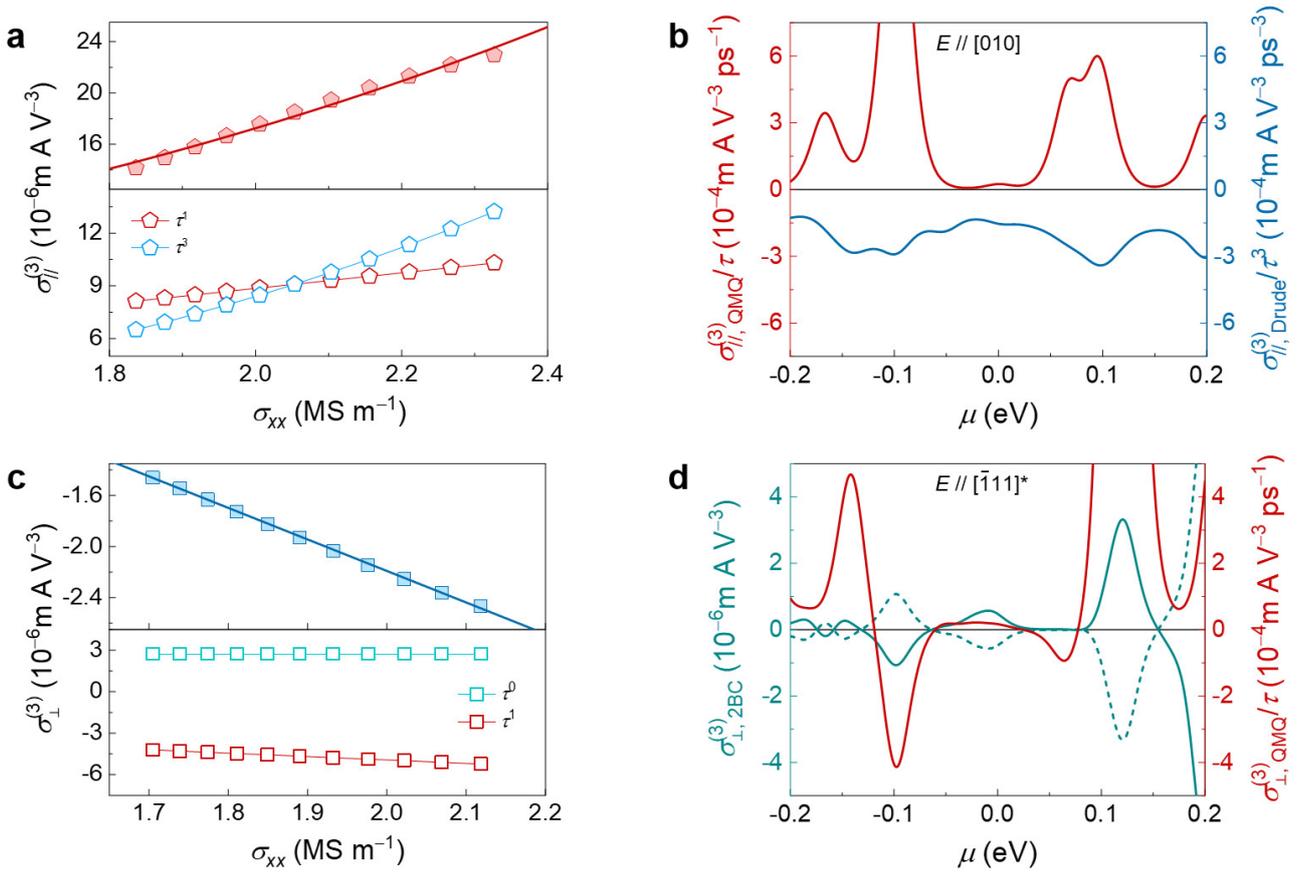

**Fig. 4 | Microscopic mechanisms of the third-order electrical transport in (101)-oriented RuO$_2$ thin films. a**, Experimentally-obtained third-order longitudinal conductivity $\sigma_{//}^{(3)}$ versus first-order longitudinal conductivity $\sigma_{xx}$ at 200–300 K (upper panel). The solid line is fitting to the data according to Eq. 4. The lower panel displays the respective contributions to $\sigma_{//}^{(3)}$ that are proportional to $\tau$ and $\tau^3$. $I_x^\omega$ is applied along [010] in the measurement. **b**, First-principles calculation results of third-order longitudinal conductivity contributed by quantum metric quadrupoles $\sigma_{//,\mathrm{QMQ}}^{(3)}$ (left axis, divided by $\tau$) and Drude-like mechanisms $\sigma_{//,\mathrm{Drude}}^{(3)}$ (right axis, divided by $\tau^3$). $\tau$ with typical value of ~0.04 ps denotes electron scattering time. $I_x^\omega$ is applied along [010] in the calculation. **c**, Experimentally-obtained third-order Hall conductivity $\sigma_\perp^{(3)}$ versus $\sigma_{xx}$ at 200–300 K (upper panel). The solid line is fitting to the data according to Eq. 5. The lower panel displays the respective contributions to $\sigma_\perp^{(3)}$ that are proportional to $\tau^0$ and $\tau^1$. $I_x^\omega$ is applied along $[\bar{1}11]^*$ in the measurement. **d**, First-principles calculation results of third-order Hall conductivity contributed by intrinsic second-order Berry curvature $\sigma_{\perp,\mathrm{2BC}}^{(3)}$ (left axis) and quantum metric quadrupoles $\sigma_{\perp,\mathrm{QMQ}}^{(3)}$ (right axis, divided by $\tau$). The dash line denotes results after $\mathcal{T}$. $I_x^\omega$ is applied along $[\bar{1}11]^*$ in the calculation.

**Methods**

**Sample preparation**

Epitaxial (101)-RuO$_2$ thin films were grown with a pulsed-laser deposition system with a base pressure of 2 × 10$^{-6}$ Pa. The detailed fabrication protocols can be found in our previous report[14]. The as-deposited samples were then processed into Hall bars arranged along RuO$_2$ [010], [$\bar{1}$11]*, or [$\bar{1}$01] with ultra-violet lithography and Ar ion milling. The widths of the current channel and the Hall branches are both 5 μm while the spacing between two Hall crosses is 25 μm.

**Electrical transport measurements**

Electrical contacts between the Hall bar electrodes and aluminum wires were made with an ultrasonic wire bonder. A Quantum Design VersaLab system was exploited to study the dc transport properties. For the nonlinear transport measurements, a Keithley 6221 source meter was employed to apply ac currents to the sample, and two Stanford SR830 lock-in amplifiers were utilized to simultaneously detect the first and second/third harmonic voltage responses. The ac current was applied along the +$x$ direction (from source to ground), and the voltage contacts were made so that a positive differential voltage corresponds to an electric field along the +$x$ or +$y$ direction. The frequency of the a.c. current is 13.000 Hz unless otherwise specified.

**Field-cooling experiments**

In the field-cooling experiments, the samples were heated to ~423 K in air in a high-magnetic-field annealing furnace. After the temperature had become stable, a magnetic field of ~30 T was applied along the out-of-plane direction of the thin films. We annealed the samples in the magnetic field for 3–5 mins, and then cooled them down to room temperature. The magnetic field was removed when the temperature reached ~323 K.

**First-principles calculations**

The first-principles calculations were performed using the Vienna *ab initio* simulation package[55] within the projector augmented wave method[56] and the generalized gradient approximation of the Perdew-Burke-Ernzerhof[57] exchange-correlation functional. The cutoff energy of 520 eV and 12 × 12 × 18 Γ-centered $k$-point meshes were used for bulk RuO$_2$. The crystal structure of RuO$_2$ was fully relaxed until the residual force of each atom is less than 0.001 eV Å$^{-1}$. The optimal lattice constants are $a = 4.544$ Å and $c = 3.14$ Å. The (101)-oriented thin films with different thickness were built by using a slab model where a vacuum thickness of 16 Å was employed. For thin films, a $10 \times 12 \times 1$ Γ-centered $k$-point mesh was used. The spin-orbit coupling effect was considered in our electronic structure calculations. The on-site Coulomb correlation about the $d$ orbitals of Ru atoms was considered within the DFT+$U$ approach[58] with the effective Hubbard parameter $U$ tested for bulk system. Through comparison with several recent studies (Supplementary Note 1), $U = 1.0$ eV was finally adopted, which results in a nonmagnetic ground state for bulk $RuO_2$ but altermagnetic states for (101) thin films (Extended Data Fig. 1). The orientation of the Néel vector is along [001] axis direction. The tight-binding Hamiltonians based on the maximally localized Wannier functions (MLWF) (ref. [59]) with Ru-$d$ and O-$p$ orbitals were constructed to further calculate various quantum geometry quantities and nonlinear transport properties by using the WannierTools package[60]. The third-order longitudinal and Hall conductivities are calculated at $T = 100$ K with more than $3000^2$ $\boldsymbol{k}$ points.

To compare with the result of scaling analysis, we calculate the $\sigma_{//}^{\text{Drude}}(\psi = 0°)$, $\sigma_{//}^{\text{QMQ}}(\psi = 0°)$, $\sigma_{\perp}^{\text{int}}(\psi = 45°)$, $\sigma_{\perp}^{\text{QMQ}}(\psi = 45°)$ contributions in a 4-layer slab model. The third-order conductivity tensor is defined as $J_\alpha^{(3)} = \sigma_{\alpha\beta\gamma\delta} E_\beta E_\gamma E_\delta$, where the Greek letters denote directions along principal axis. When the injected current is at an angle $\psi$ to the $b$ axis, the relation between the measured $\sigma_{//}^{3\omega}$, $\sigma_{\perp}^{3\omega}$ with the third-order conductivity tensor can be derived as

$$\sigma_{//}^{(3)}(\Psi) = \frac{\sigma_1^{(3)}\rho_b^3 \cos^4\Psi + \sigma_2^{(3)}\rho_a^3 \sin^4\Psi + 3\left[\sigma_3^{(3)}\rho_b\rho_a^2 + \sigma_4^{(3)}\rho_b^2\rho_a\right]\cos^2\Psi \sin^2\Psi}{(\rho_b \cos^2\Psi + \rho_a \sin^2\Psi)^3} \tag{6}$$

$$\sigma_{\perp}^{(3)}(\Psi) = \frac{\left[3\sigma_4^{(3)}\rho_b^2\rho_a - \sigma_1^{(3)}\rho_b^3\right]\cos^3\Psi \sin\Psi + \left[\sigma_2^{(3)}\rho_a^3 - 3\sigma_3^{(3)}\rho_b\rho_a^2\right]\cos\Psi \sin^3\Psi}{(\rho_b \cos^2\Psi + \rho_a \sin^2\Psi)^3} \tag{7}$$

where $\sigma_1^{(3)} = \sigma_{bbbb}$, $\sigma_2^{(3)} = \sigma_{aaaa}$, $\sigma_3^{(3)} = \frac{1}{3}(\sigma_{bbaa} + \sigma_{baba} + \sigma_{baab})$ and $\sigma_4^{(3)} = \frac{1}{3}(\sigma_{aabb} + \sigma_{abab} + \sigma_{abba})$. Here $a, b$ axes of the thin film correspond to $[\bar{1}01]$ and $[010]$ directions of bulk $RuO_2$, respectively. The other in-plane tensor elements vanish due to the glide mirror plane $\widetilde{\mathcal{M}}_{010}$. The Drude-like, QMQ and intrinsic third-order effects are calculated using the following formulas[18,38]

$$\sigma_{\alpha\beta\gamma\delta}^{\text{Drude}} = \tau^3 \frac{e^4}{4\hbar^4} \sum_n \int_{\boldsymbol{k}} f_0 \partial_\alpha \partial_\beta \partial_\gamma \partial_\delta \varepsilon_{n,\boldsymbol{k}}$$

$$= -\tau^3 \frac{e^4}{4\hbar^4} \sum_n \int_{\boldsymbol{k}} v_{n,\boldsymbol{k}}^\alpha f_0' \partial_\beta \partial_\gamma \partial_\delta \varepsilon_{n,\boldsymbol{k}} \tag{8}$$

$$\sigma_{\alpha\beta\gamma\delta}^{\text{QMQ}} = \tau \frac{e^4}{4\hbar^2} \sum_n \int_{\mathbf{k}} \left[ 2\left( -\partial_\alpha \partial_\beta G_{n,\mathbf{k}}^{\gamma\delta} + \partial_\alpha \partial_\delta G_{n,\mathbf{k}}^{\beta\gamma} - \partial_\beta \partial_\delta G_{n,\mathbf{k}}^{\alpha\gamma} \right) f_0 + v_{n,\mathbf{k}}^\alpha v_{n,\mathbf{k}}^\beta G_{n,\mathbf{k}}^{\gamma\delta} f_0'' \right]$$

$$= \tau \frac{e^4}{4\hbar^2} \sum_n \int_{\mathbf{k}} \left[ \partial_\beta G_{n,\mathbf{k}}^{\gamma\delta} v_{n,\mathbf{k}}^\alpha - 2\left( \partial_\alpha G_{n,\mathbf{k}}^{\beta\gamma} - \partial_\beta G_{n,\mathbf{k}}^{\alpha\gamma} \right) v_{n,\mathbf{k}}^\delta - \left( \partial_\alpha \partial_\beta \varepsilon_{n,\mathbf{k}} \right) G_{n,\mathbf{k}}^{\gamma\delta} \right] f_0' \quad (9)$$

$$\sigma_{\alpha\beta\gamma\delta}^{\text{int}} = \frac{e^4}{4\hbar} \sum_n \int_{\mathbf{k}} f_0 \left[ \partial_\beta T_{n,\mathbf{k}}^{\alpha\gamma\delta} - \partial_\alpha T_{n,\mathbf{k}}^{\beta\gamma\delta} \right]$$

$$= \frac{e^4}{4\hbar} \sum_n \int_{\mathbf{k}} \left[ v_{n,\mathbf{k}}^\alpha T_{n,\mathbf{k}}^{\beta\gamma\delta} - v_{n,\mathbf{k}}^\beta T_{n,\mathbf{k}}^{\alpha\gamma\delta} \right] f_0' \quad (10)$$

Here $G_{n,\mathbf{k}}^{\alpha\beta} = \text{Re} \sum_{m \neq n} \mathcal{A}_{nm,\mathbf{k}}^\alpha \mathcal{A}_{mn,\mathbf{k}}^\beta / (\varepsilon_{n,\mathbf{k}} - \varepsilon_{m,\mathbf{k}})$ is the band-resolved Quantum metric, where $\mathcal{A}_{n,\mathbf{k}}^a = i\langle u_{n,\mathbf{k}} | \partial_a u_{n,\mathbf{k}} \rangle$ is the Berry connection, $m$, $n$ denote bands not degenerate with each other, $\partial_a$ is short for $\partial/\partial k_a$, $v_{n,\mathbf{k}}^\alpha \equiv \partial_\alpha \varepsilon_{n,\mathbf{k}}$, and $f_0$ is the Dirac-Fermi distribution function. $T_{n,\mathbf{k}}^{\alpha\beta\gamma}$ is the second-order Berry-connection polarizability tensor, and its detailed definition can be seen in ref.[18]. Since the frequency dependence is negligible in the experimental results, $\omega\tau \approx 0$ is taken. The second line of Eqs. 8–10 is the form of integration by parts, which is utilized in the calculations. Since the observed anisotropy $\rho_b/\rho_a$ is small, we assume $\rho_b \approx \rho_a$ for simplification. Then we rotate the Cartesian coordinate system by an angle $\Psi$ in-plane in the calculation of four-layer RuO$_2$ slab, which is equivalent to using Eqs. 6 and 7.

## Data availability

The data that support the findings of this study are available within the Article and its Supplementary Information. Further information is available from the corresponding author on reasonable request.

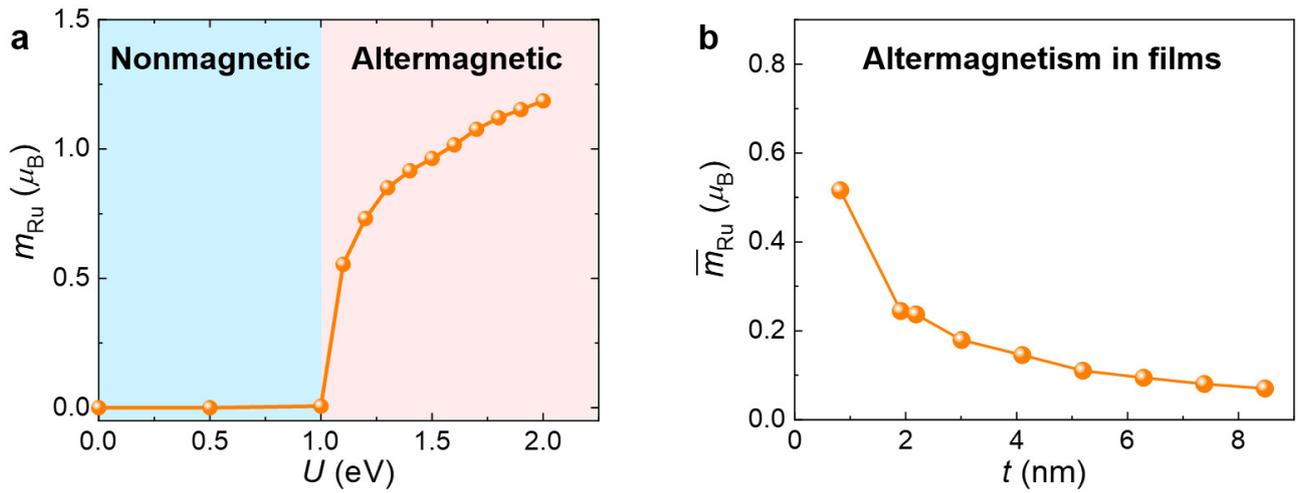

**Extended Data Fig. 1 | Calculated magnetic properties of bulk RuO$_2$ and (101)-oriented thin films. a**, Magnetic moment of each Ru atom $m_{Ru}$ as a function of the effective Hubbard parameter $U$ for a bulk system, where the nonmagnetic and altermagnetic regions are highlighted in blue and red, respectively. **b**, Average magnetic moment of each Ru atom $\bar{m}_{Ru}$ as a function of layer thickness $t$ of (101) thin films, wherein $U = 1$ eV is employed. We mainly study the nonlinear transport in samples of 5–9 nm, wherein the average magnetic moment exhibits a small variation with $t$.

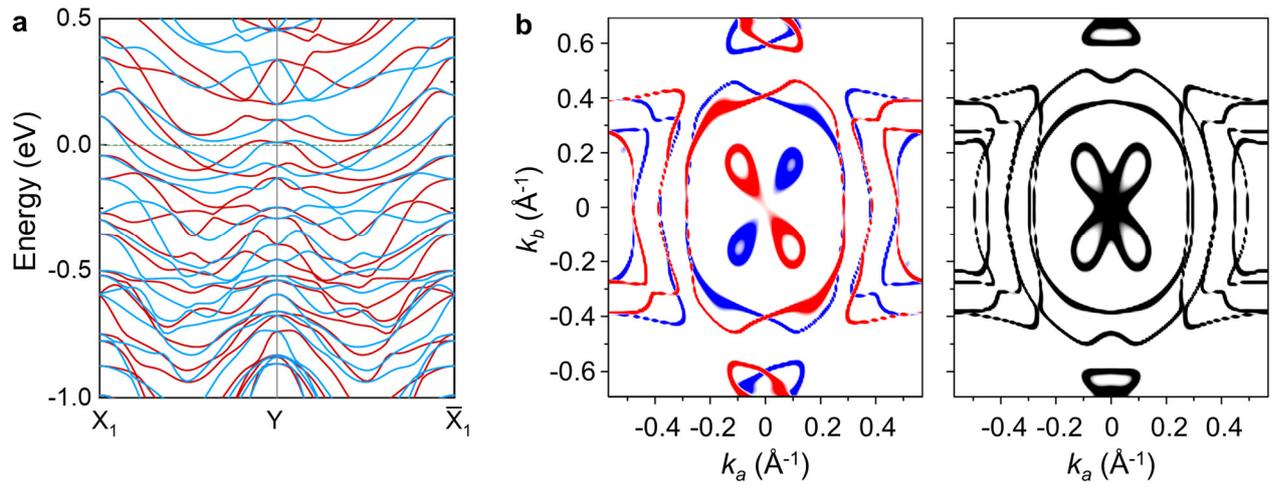

**Extended Data Fig. 2 | Spin-split electronic bands of a four-layer (101)-oriented thin slab. a**, Calculated electronic band structures in the absence of spin-orbit coupling, where $U = 1$ eV is employed. Blue and red bands are of electrons with opposite spin. **b**, Calculated Fermi surfaces without (left) and with (right) spin-orbit coupling. Blue and red lines are of electrons with opposite spin.

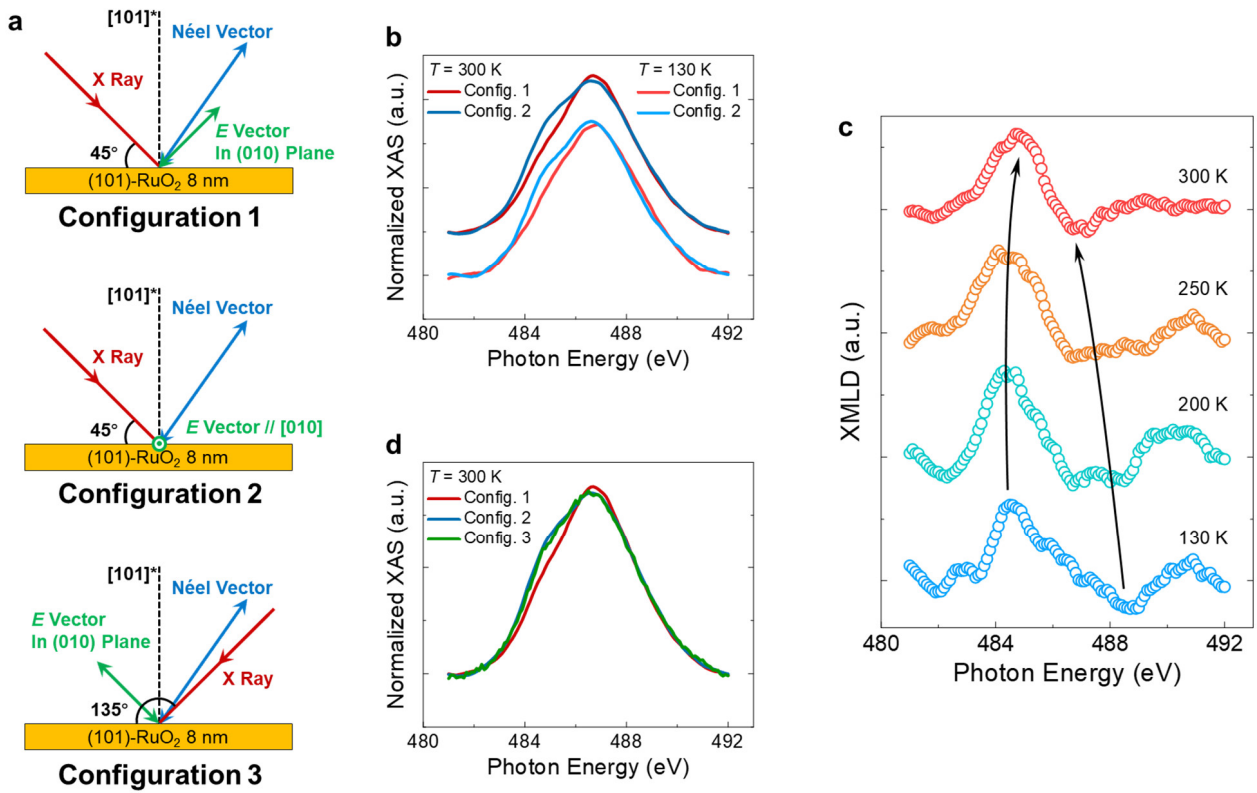

**Extended Data Fig. 3 | X-ray magnetic linear dichroism (XMLD) in an 8-nm-thick (101)-oriented $RuO_2$ thin film. a**, Schematic illustration on the measurement geometry. In configuration 1, the electric field vector ***E*** of the linearly polarized incident x ray is nearly parallel to the Néel vector of $RuO_2$. In configurations 2 and 3, ***E*** is (nearly) perpendicular to the Néel vector. **b**, Normalized x-ray absorption (XAS) spectra measured in configurations 1 and 2 at a temperature $T$ of 300 and 130 K. **c**, $T$-dependent differential XMLD spectra deduced from the data measured in configurations 1 and 2. The arrows highlight the shift of peak positions with $T$. **d**, Normalized XAS spectra measured in configurations 1–3 at room temperature.

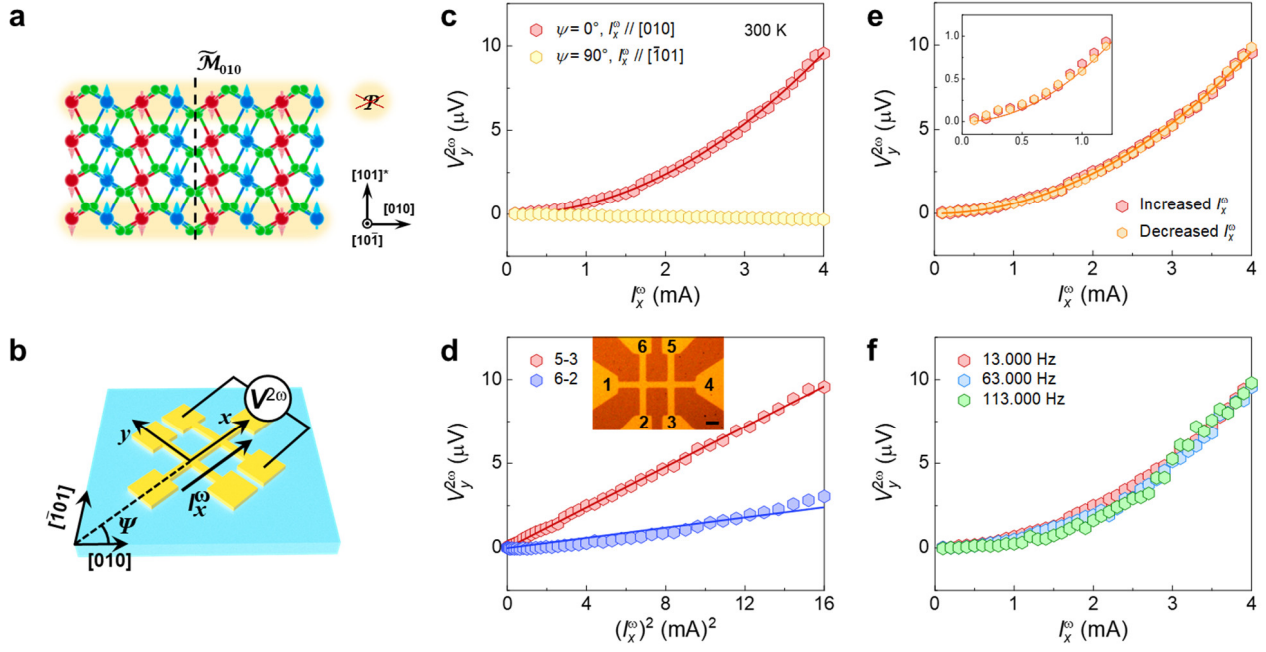

**Extended Data Fig. 4 | Second-order Hall effect induced by interface/surface symmetry breaking in an 8-nm-thick (101)-oriented RuO$_2$ thin film. a**, Side view of a (101)-RuO$_2$ thin film, which possesses a glide mirror plane $\widetilde{\mathcal{M}}_{010}$. The parity ($\mathcal{P}$) symmetry is additionally broken at surface and interface. Due to the tetragonal structure of RuO$_2$, [101]$^*$ is slightly tilted away from the [101] direction. **b**, Schematic illustration on the measurement protocol. $\Psi$ denotes the angle between the $x$ axis of the coordinate system and the [010] direction. **c**, Second-harmonic transverse voltages $V_y^{2\omega}$ as a function of an ac current $I_x^\omega$ applied along [010] and [$\bar{1}$01] at room temperature. **d–f**, $V_y^{2\omega}(I_x^\omega)$ relations measured in different conditions. $I_x^\omega$ is applied along [010]. The inset in **d** displays the optical image of a typical device wherein the scale bar denotes a length of 5 μm. The electrodes are numbered to indicate which pair is employed in the measurement. The inset in **e** displays the data collected with a small $I_x^\omega$. The solid lines in **c** and **e** are square fittings to the data. The solid lines in **d** are linear fittings to the data.

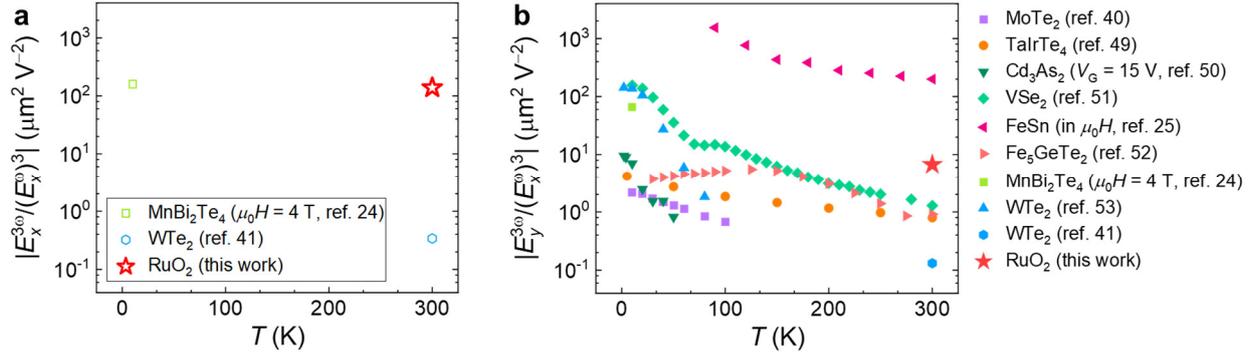

**Extended Data Fig. 5 | Giant magnitude of the room-temperature third-order electrical transport in (101)-oriented RuO$_2$ thin films. a** and **b**, Comparison of $E_x^{3\omega}/(E_x^{\omega})^3$ and $E_y^{3\omega}/(E_x^{\omega})^3$ at different temperature $T$ in various materials with those of an 8-nm-thick (101)-RuO$_2$ thin film (the twelve-terminal device). Note that FeSn and MnBi$_2$Te$_4$ exhibit large third-order responses in the presence of large magnetic fields $\mu_0H$, while the $E_y^{3\omega}/(E_x^{\omega})^3$ of Cd$_3$As$_2$ was collected under finite gating voltage $V_G$. The data points are reproduced from refs. [25,40,49–53] and estimated from the results in refs. [24,41].

# Contents

**Supplementary Notes**

**Note 1: Brief review on the altermagnetism of $RuO_2$**

The antiferromagnetism of $RuO_2$ was initially demonstrated in two early works in bulk single crystals[1,2] and thin films[2]. Nevertheless, many recent theoretical and experimental studies question the conclusions and report the absence of long-range magnetic order in $RuO_2$ (refs. [3–11]), making the antiferromagnetism of $RuO_2$ currently under debate. Here we briefly review the latest research progress and present our understanding on this issue.

(1) Ground state of *bulk* $RuO_2$

Latest theoretical studies[6,12] indicated the fragility of the envisaged long-range altermagnetic order of bulk $RuO_2$, possibly due to its proximity to the Landau-Pomeranchuk instability. In particular, it was shown that, in order to arrive at an altermagnetic ground state, a rather large Hubbard $U$ of at least ~1.23 eV is indispensable in first principles calculations, unless imperfections such as proper hole doping and (or) strain are incorporated.

Regarding experimental investigations, the absence of long-range magnetic order in bulk single-crystal $RuO_2$ has been demonstrated by accumulating experimental studies[3–5,7,9,10]. However, there remains one exception[13], wherein the altermagnetic *d*-wave spin texture in momentum space was revealed by using spin- and angle-resolved photoemission spectroscopy. We note that the overall consistency of the conclusions drawn by independent experimental groups, *i.e.*, a nonmagnetic ground state of bulk $RuO_2$, seems also in line with its proposed fragile altermagnetism—the bulk single crystals employed in different studies are generally of high quality and, consequently, tend to be nonmagnetic.

Therefore, we may infer that a high-quality bulk $RuO_2$ single crystal is likely, or at least in most cases, nonmagnetic.

(2) Ground state of *thin-film* $RuO_2$

For the thin films of $RuO_2$, the situation is much more complicated. This is not only due to the technical challenges in characterizing the magnetic properties of thin films, but also rooted in the variety and complexity of the defects in correlated oxide-thin-film materials, including but not limited to doping/stoichiometry, surface/interface, and epitaxial strain, which can significantly influence the properties of RuO$_2$ with a fragile altermagnetic ground state. For instance, it has been theoretically demonstrated that the envisaged altermagnetic ground state of RuO$_2$ can be stabilized by hole doping and (or) epitaxial strain without the necessity of a large Hubbard $U$ (refs. [6,12,14]). Moreover, according to the calculation results in ref. [15], altermagnetism even naturally emerges in (001)-oriented thin films as a result of strong interlayer relaxation that completely takes the place of Hubbard $U$ correction.

Consistent with the intricateness, efforts made by different groups to experimentally ascertain the magnetic ground state in RuO$_2$ thin films may lead to conflicting results even with the same characterization techniques. For instance, ref. [4] claimed no magnetism was found with $\mu$SR in a 11-nm-thick film, while a very recent work[16] exploited low-energy $\mu$SR and better depth resolution to confirm the existence of surface (~10 nm) magnetism in 30-nm-thick RuO$_2$ samples; the temperature dependence of x-ray magnetic linear dichroism that can indicate altermagnetic order was reported to be absent in ref. [11] but present and prominent in refs. [17,18], and refs. [19,20] further revealed the Néel-vector dependence. In addition, a series of latest studies have convincingly demonstrated the existence of altermagnetism in RuO$_2$ thin films based on various other magnetic-order probes. Specifically, ref. [21] directly revealed the time-reversal symmetry breaking in momentum space due to altermagnetism of 34-nm-thick thin films by using magnetic circular dichroism (MCD) in angle-resolved photoemission spectroscopy, and ref. [22] further detected MCD in core-level x-ray photoelectron spectroscopy of 7-nm-thick samples; refs. [23,24] corroborated the altermagnetism of ~10-nm-thick thin films by observing polarity-reversable tunneling magnetoresistance and inverse spin-splitting effect through flipping the Néel vector of RuO$_2$; ref. [25] solidified the altermagnetism of 10- and 20-nm thick samples by detecting the room-temperature tunneling anisotropic magnetoresistance effect induced by the spin-flop transition in RuO$_2$/MgO/RuO$_2$ tunnel junctions; ref. [26] proved the altermagnetism in at least the near-surface regions of 60-nm-thick samples by using point-contact Andreev reflection spectroscopy. We also notice that the existence of altermagnetism indeed seems to intimately relate to defects such as strain and surface/interface symmetry breaking[16,22,26,27], and that all of the aforementioned studies indicated an altermagnetic order consistent with the early report on RuO$_2$ thin films[2], *i.e.*, a Néel temperature above 300 K (ranging from 390 to 450 K in the aforementioned studies) and a Néel vector aligned (nearly) parallel to the *c* axis.

Therefore, we conclude that it is practical to realize the envisaged altermagnetic order in RuO$_2$ *thin films*. In addition, the severe controversy among different experimental studies could result from a simple fact that the RuO$_2$ thin films under independent investigation possess varied types and extents of defects and are, consequently, distinct materials.

**Note 2: Evidence of altermagnetism in (101)-RuO$_2$ thin films**

As illustrated in Fig. S1b, the Hall resistivity of an 8 nm-thick sample scales linearly with the applied out-of-plane magnetic field, which excludes ferromagnetism in the RuO$_2$ thin film. On the other hand, compared to a single layer, the Co layer in adjacent to a 20-nm-thick (101)-RuO$_2$ thin film exhibits a shifted magnetic hysteresis loop and an enhanced magnetic coercivity after field cooling from 395 to 50 K (Fig. 2c in [*Adv. Mater.* **37**, 2507764 (2025)])—characteristics of the exchange bias effect between ferromagnets and antiferromagnets[28]. This clearly supports the presence of antiferromagnetism in the RuO$_2$ layer. We define the effective exchange-bias field as $\mu_0 H_{EB} = (\mu_0 H_{C1} + \mu_0 H_{C2})/2$ and the magnetic coercivity of Co as $\mu_0 H_C = (\mu_0 H_{C1} - \mu_0 H_{C2})/2$, wherein $\mu_0 H_{C1}$ and $\mu_0 H_{C2}$ are the positive and negative magnetic coercivity, respectively. Notably, the $\mu_0 H_{EB}$ and $\mu_0 H_C$ are both higher for the exchange bias established by field cooling along RuO$_2$[$\bar{1}$01] than along RuO$_2$[010]. This implies a larger in-plane projection of the Néel vector along [$\bar{1}$01], consistent with the *c*-axis-oriented Néel vector in RuO$_2$ thin films reported in previous studies[2,17–26].

We next further corroborate the altermagnetism in our samples with temperature- and angle-dependent x-ray absorption spectroscopy (XAS). For linearly-polarized x ray, the line shape and intensity of the absorption spectra of an antiferromagnet is modified by the relative orientation $\theta$ of the polarization direction with respect to the Néel vector, with the difference between $\theta = 0°$ and $\theta = 90°$ defining the x-ray magnetic linear dichroism (XMLD). Compared to conventional magnetic characterizations, XAS and XMLD is particularly sensitive to surface ($10^0$–$10^1$ nm scale) antiferromagnetism, and is thus applicable to the 8-nm-thick samples mainly studied in this work.

Given that the electrical transport properties reveal the existence of a glide mirror plane perpendicular to RuO$_2$[010], the Néel vector of the samples, if present, will be constrained in the (010) plane. Consequently, we first examined the XAS of an 8-nm-thick thin film with the incident x ray polarized parallel and perpendicular to RuO$_2$(010), *i.e.*, configurations 1 and 2 in Extended Data Fig. 3a. As depicted in Extended Data Fig. 3b, the distinction between the Ru $M_2$ ($3p_{1/2}$) absorption edge in the two spectra is notable at room temperature, and the corresponding differential dichroism is displayed in Extended Data Fig. 3c. Note that in addition to antiferromagnetism, structural asymmetry may also contribute to XAS dichroism, *i.e.*, x-ray linear dichroism (XLD), since XAS also incorporates information on the Ru orbital orientation. In order to shed light on the magnetic origin of the dichroism, we subsequently performed temperature-dependent measurements. As shown in Extended Data Fig. 3b, decreasing temperature from 300 to 130 K results in notable changes in the line shape of the XAS curves, which is captured in the differential dichroism in a clearer manner (Extended Data Fig. 3c). Moreover, the peak positions in the spectra gradual shift and the dichroism strength tends to attenuate with lowered temperature. Such systematic temperature dependence of XAS corresponds to either a structural phase change or suppressed antiferromagnetism in the sample at elevated temperature, because XLD induced by structural asymmetry is temperature-independent, especially in the investigated narrow temperature range. Given that there is no structural transition of both $RuO_2$ and $TiO_2$ at 130–300 K according to the smooth $\rho_{xx}(T)$ and $d\rho_{xx}/dT(T)$ curves in Fig. S1a, the prominent temperature dependence of XAS can be exclusively ascribed to the presence of antiferromagnetism in the sample. We also notice that similar temperature-dependent XMLD results have been recently reported for (101), (100), and (110)-$RuO_2$ thin films[17,18].

We subsequently try to ascertain the Néel-vector orientation of the 8-nm-thick sample by angle-dependent XAS measurement. Since both previous studies and our exchange-bias measurements indicate a Néel vector along the $c$ axis in $RuO_2$ thin films, we altered the incident angle of the x ray from 45° in configuration 1 to 135° in configuration 3 (Extended Data Fig. 3a) and studied the resultant variation in XAS. For a $c$-axis-oriented Néel vector, this corresponds to a change of $\theta$ from 10° (polarization nearly parallel to the Néel vector) to 80° (polarization nearly perpendicular to the Néel vector). As shown in Extended Data Fig. 3d, the XAS spectrum collected in configuration 3 almost overlaps with that in configuration 2, implying that the polarization is practically normal to the Néel vector in this case. Therefore, we conclude that the Néel vector of the sample is along the $c$ axis of $RuO_2$. We notice that such an antiferromagnetic order is in line with other recent experimental results, and is also the one widely exploited in previous studies[2,17–26] on the altermagnetism of $RuO_2$.

We finally solidify the above conclusion with density functional theory calculations. Consistent with the previous calculations[29], the magnetic ground state of bulk $RuO_2$ strongly depends on the value of $U$, as shown in Extended Data Fig. 1a. Based on our DFT calculations, when using $U = 2$ eV, $RuO_2$ is an altermagnet with a magnetic anisotropy energy of 6 meV, *i.e.*, the Néel vector along the [001] direction is 6 meV lower in energy than along the [110] direction. However, when $U = 1$ eV is adopted, the ground state of bulk becomes non-magnetic. It is noteworthy that the (101)-oriented $RuO_2$ thin film exhibits an altermagnetic ground state even at the same $U = 1$ eV. For instance, the spin-split electronic band structures and Fermi surfaces of a four-layer slab are illustrated in Extended Data Fig. 2. As shown in Extended Data Fig. 1b, with $U = 1$ eV, the average magnetic moment of each Ru atom gradually decreases as the layer thickness increases; however, for experimentally fabricated films ranging from 3 nm to 8 nm, the average magnetic moment shows minimal variation. Importantly, the calculated Néel vector of the thin film spontaneously converges to the [001] crystal axis direction.

**Note 3: Symmetry analyses on the third-order electrical transport in (101)-RuO$_2$ thin films**

As detailed in Table S1, rutile-structured (101)-oriented thin films preserve the symmetry of inversion $\mathcal{P}$, time reversal $\mathcal{T}$, glide mirror $\widetilde{\mathcal{M}}_{010} = \{\mathcal{M}_{010}|1/2,1/2,0\}$, two-fold screw rotation $\widetilde{C}_{010}^2 = \{C_{010}^2|1/2,1/2,0\}$, and their respective combination with $\mathcal{T}$ (here the main axes $a$, $b$, and $c$ for the thin films are defined along $[\bar{1}01]$, $[010]$, and $[101]^*$ directions of bulk RuO$_2$, respectively). The existence of collinear antiferromagnetic order cannot affect $\mathcal{P}$, but would violate $\mathcal{T}$ and $\mathcal{P}\mathcal{T}$. On the other hand, whether $\widetilde{\mathcal{M}}_{010}(\mathcal{T})$ and $\widetilde{C}_{010}^2(\mathcal{T})$ survive the emergence of altermagnetism is dictated by the orientation of Néel vector $\boldsymbol{N}$. Note that the presence of $\mathcal{P}\mathcal{T}$ forbids any third-order $\mathcal{T}$-odd effect, while $\widetilde{\mathcal{M}}_{010}$ strictly restricts the third-order Hall effect when the current is injected within or perpendicular to the (010) glide mirror plane. Therefore, as detailed in Table S2, the existence of altermagnetic order in the samples would result in third-order $\mathcal{T}$-odd responses, while the direction of $\boldsymbol{N}$ determines the crystalline anisotropy of the third-order Hall effect.

**Table S1 | Symmetry of (101)-RuO$_2$ thin films in different magnetic states.** NM and AM stand for nonmagnetic and altermagnetic, respectively.

| Magnetic state of (101)-RuO$_2$ | $\mathcal{P}$ | $\mathcal{P}\mathcal{T}$ | $\widetilde{\mathcal{M}}_{010}$, $\widetilde{C}_{010}^2$ | $\widetilde{\mathcal{M}}_{010}\mathcal{T}$, $\widetilde{C}_{010}^2\mathcal{T}$ |
|---|---|---|---|---|
| NM | √ | √ | √ | √ |
| AM $\boldsymbol{N}\,//\,[001]$ | √ | × | √ | × |
| AM $\boldsymbol{N}\,//\,[101]$ | √ | × | √ | × |
| AM $\boldsymbol{N} \not\subset (010)$ and $\boldsymbol{N} \not\!/\![010]$ | √ | × | × | × |
| AM $\boldsymbol{N}\,//\,[010]$ | √ | × | × | √ |

**Table S2 | Symmetry constraints on the third-order conductivities in (101)-RuO$_2$ thin films in different magnetic states.** $\Psi$ denotes the angle between the applied electric field $E_{//}^{\omega}$ and the [010] direction. $n$ is an arbitrary integer. $\Psi = n\pi/2$ indicates that $E_{//}^{\omega}$ is applied within or perpendicular to the (010) plane. QMQ, BCQ, and Int represent quantum metric quadrupole, Berry curvature quadrupole, and intrinsic (which results in electric-field-induced second-order Berry curvature) mechanisms of third-order longitudinal (transverse) conductivity $\sigma_{//}^{(3)}$ ($\sigma_{\perp}^{(3)}$), respectively. The nonlinear electrical transport properties of the sample are consistent with the row highlighted in red.

| Magnetic state of (101)-RuO$_2$ | $E_{//}^{\omega}$ relative to [010] | $\sigma_{//, \text{QMQ}}^{(3)}$ | $\sigma_{//, \text{Drude}}^{(3)}$ | $\sigma_{\perp, \text{Int}}^{(3)}$ | $\sigma_{\perp, \text{QMQ}}^{(3)}$ | $\sigma_{\perp, \text{BCQ}}^{(3)}$ |
|---|---|---|---|---|---|---|
| NM | $\Psi = n\pi/2$ | √ | √ | × | × | × |
| NM | $\Psi \neq n\pi/2$ | √ | √ | × | √ | × |
| <span style="color:red">AM</span> <span style="color:red">$N \subset (010)$</span> | <span style="color:red">$\Psi = n\pi/2$</span> | <span style="color:red">√</span> | <span style="color:red">√</span> | <span style="color:red">×</span> | <span style="color:red">×</span> | <span style="color:red">×</span> |
| <span style="color:red">AM</span> <span style="color:red">$N \subset (010)$</span> | <span style="color:red">$\Psi \neq n\pi/2$</span> | <span style="color:red">√</span> | <span style="color:red">√</span> | <span style="color:red">√</span> | <span style="color:red">√</span> | <span style="color:red">√</span> |
| AM $N \not\subset (010)$ and $N \not\parallel$ [010] | Any $\Psi$ | √ | √ | √ | √ | √ |
| AM $N // $ [010] | $\Psi = n\pi/2$ | √ | √ | √ | × | √ |
| AM $N // $ [010] | $\Psi \neq n\pi/2$ | √ | √ | √ | √ | √ |

We found that the third-order Hall effect reverses sign after flipping the altermagnetic order (Fig. 2h in the main text). This demonstrates that the measured third-order Hall conductivity $\sigma_{\perp}^{(3)}$ is dominated by $\mathcal{T}$-odd mechanisms such as intrinsic second-order field-induced Berry curvature and Berry curvature quadruples, thus corroborating the presence of altermagnetism in the sample. Furthermore, as displayed in Fig. 3e in the main text, the crystalline anisotropy of the third-order Hall effect exhibits prominent two-fold symmetry; $\sigma_{\perp}^{(3)}$ almost vanishes when the current is applied within or perpendicular to the (010) plane. These features reveal the existence of a (glide) mirror plane in (010), indicating that $N$ is oriented within the (010) plane in the sample. Therefore, the nonlinear electrical transport properties of the sample are well consistent with the altermagnetic order revealed by the exchange-bias and x-ray magnetic linear dichroism measurements.

**Note 4: Second-order Hall effect in (101)-RuO$_2$ thin films**

As illustrated in Extended Data Fig. 4a, parity ($\mathcal{P}$) symmetry is extrinsically violated at the surface and interface of RuO$_2$ thin films so that $\mathcal{T}$-odd quantum metric dipoles (QMDs) and $\mathcal{T}$-even Berry curvature dipoles (BCDs) can be allowed. Thus, we detect the second-order Hall effect to unveil the existence of these quantum geometric quantities. We measure the amplitude of the second-harmonic Hall voltages $V_y^{2\omega}$ generated by an applied ac current of $I_x^{\omega}\sin(\omega t)$, wherein $I_x^{\omega}$ and $\omega$ are the amplitude and angular frequency, respectively (Extended Data Fig. 4b). Note that the $\widetilde{\mathcal{M}}_{010}$ in (101)-RuO$_2$ thin films forbids any second-order Hall response when $I_x^{\omega}$ is applied parallel to it[30,31]. Accordingly, as displayed in Extended Data Fig. 4c, $V_y^{2\omega}$ almost vanishes for $I_x^{\omega}$ along [$\bar{1}$01], but is finite and proportional to the square of $I_x^{\omega}$ for $I_x^{\omega}$ along [010]. We thus focus on the latter case in subsequent studies. We confirm that the nonzero $V_y^{2\omega}$ is irrelevant to extrinsic artifacts such as the thermoelectric effect, contact junctions, and capacitive coupling, since the $V_y^{2\omega}(I_x^{\omega})$ relation exhibits negligible dependence on the measurement protocols as shown in Extended Data Fig. 4e,f.

Note that despite the second-order nature, $V_y^{2\omega}$ is significantly lower than the third-harmonic transverse voltage generated by the same $I_x^{\omega}$ (Fig. 2f in the main text), which is suggestive of an extrinsic effect. Further evidence for the interface/surface origin of the second-order Hall effect can be found in the thickness dependence. As shown in Fig. S6d, the second-order response strongly depends on thickness, and is much more pronounced in thinner samples, while the third-order effects are almost independent on thickness. This demonstrates that the second-order transverse transport is an interfacial effect, while the third-order transport is of bulk derivation. Notably, the magnitudes of $V_y^{2\omega}$ are distinct for the two Hall crossings in the same device (Extended Data Fig. 4d), similar to the third-order transverse response in the main text but in sharp contrast to the case reported for $\mathcal{T}$-even materials[31]. We ascribe this feature to the existence of $\mathcal{T}$-odd mechanisms of the second-order Hall effect and the distinct altermagnetic domain population and (or) distribution in the two crossings. We also notice that the recorded $V_y^{2\omega}$ stochastically changes sign in samples with distinct thickness (Fig.

S6d), which further suggests its $\mathcal{T}$-odd characteristic.

Interestingly, the nonlinearity of the $V_y^{2\omega}(I_x^\omega)$ relation enhances with raised temperature (Fig. S5a,b), which could result from the pronounced electron scattering at high temperature. In order to make a comparison with other materials, it is convenient to deduce the second-order Hall conductivity $\sigma_\perp^{(2)}$, the calculation methods of which are detailed in Supplementary Note 6. We find that $\sigma_\perp^{(2)}$ increases to ~1.11 mA V$^{-2}$ at room temperature, which is comparable to those for some two-dimensional systems at cryogenic temperatures[30,32,33]. This indicates that the surface/interface $\mathcal{P}$ symmetry breaking in thin films is effective in giving rise to second-order Hall responses, and highlights the potential of altermagnetic thin films for energy harvesting applications such as wireless rectification[34].

We finally try to shed light on the microscopic mechanisms of the second-order Hall effect with scaling analyses. As shown in Fig. S7a, the Hall carrier density of a typical sample is less dependent on temperature at 200–300 K. We thus employ the nonlinear transport data collected in this temperature range to study the relation between $\sigma_\perp^{(2)}$ and the first-order longitudinal conductivity $\sigma_{xx}$ (Fig. S7b). For the second-order Hall effect, the QMD, BCD, Drude, and skew-scattering contributions to $\sigma_\perp^{(2)}$ are proportional to $\tau^0$, $\tau^1$, $\tau^2$, and $\tau^3$, respectively[30,31,33,35–38], wherein $\tau$ is the relaxation time of electrons. The calculation method of $\sigma_\perp^{(2)}$ is detailed in Supplementary Note 6. Since we can assume $\sigma_{xx} \propto \tau$ at 200–300 K, we employ

$$\sigma_\perp^{(2)} = C_1 + C_2\sigma_{xx} + C_3\sigma_{xx}^2 + C_4\sigma_{xx}^3 \tag{S1}$$

to fit the $\sigma_\perp^{(2)}(\sigma_{xx})$ relation measured with $I_x^\omega$ along [010] and electrodes 5-3, wherein $C_1$–$C_4$ are fitting constants. As depicted in Fig. S8a, the $\sigma_\perp^{(2)}(\sigma_{xx})$ curve can be well fitted with Eq. S1. According to the fitting results displayed in Fig. S8b, $C_2\sigma_{xx}$ and $C_3\sigma_{xx}^2$ constitute the leading contributions to $\sigma_\perp^{(2)}$ at room temperature. Note that $C_1$ and $C_3\sigma_{xx}^2$ are $\mathcal{T}$-odd contributions that can account for the observed $\mathcal{T}$-odd feature of the second-order Hall effect. In particular, $\mathcal{T}$-odd QMDs (captured by $C_1$) have been proposed to emerge in $d$-wave altermagnetic thin films due to surface/interface symmetry lowering[39], which is consistent with our results. Therefore, we may conclude that the second-order Hall effect are mainly induced by BCDs and the Drude mechanism, and the $\mathcal{T}$-odd feature are related to the latter and QMDs. In addition, we note that although most previous studies have assumed one-to-one correspondence between the four fitting terms in Eq. S1 and four distinct microscopic mechanisms, complex extrinsic scattering effects may participate in all the fitting elements[40,41]. Moreover, altermagnets with spin-split electronic bands may permit additional spin-dependent scattering channels. Therefore, we emphasize that the contributions from extrinsic scattering mechanisms may be non-negligible in our sample, the quantitative determination of which is challenging but deserves further investigation.

**Note 5: Exclusion of alternative mechanisms of the nonlinear electrical transport of $RuO_2$**

*__Contact junctions__*: The accidental formation of contact junctions between the aluminum wires and $RuO_2$ electrodes can lead to a nonlinear transport effect. In our study, this possibility can be excluded according to the following facts.

(1) The third-order longitudinal transport data measured from different pairs of electrodes are almost identical.

(2) Linear two-probe dc *I*(*V*) relations without a rectification effect were detected between any two electrodes of the samples, as shown in Fig. S2.

(3) Anisotropy of the third-order transport are consistent with the crystalline symmetry of (101)-$RuO_2$ thin films.

*__Asymmetric sample shapes__*: A globally asymmetric sample shape can result in extrinsic nonlinear transport phenomena due to the asymmetric collision of carriers against the sample boundaries. This cannot contribute to the nonlinear transport effect in our study owing to the following aspects.

(1) The Hall bars are symmetric with respect to both measurement directions, *i.e.*, the *x* and *y* axes.

(2) The second- and third-order nonlinear Hall effect exhibit remarkable crystalline anisotropy despite the almost identical shape of different Hall bars.

*__Thermoelectric effect__*: The Joule heating can give rise to a temperature gradient in the sample that would lead to nonlinear transport effects only in concert with external asymmetries such as contact junctions and asymmetric sample shapes. Therefore, it is irrelevant to the nonlinear transport of our samples.

*__Capacitive coupling__*: This can be excluded by the negligible frequency dependence of the nonlinear transport properties.

*__Resistivity variation due to Joule heating__*: The Joule heating can lead to a rise in temperature ($\propto I^2R$) and thus a change of the sample resistance, which could result in a third-order effect. This possibility can be excluded in our study according to the following aspects.

(1) The $V^{3\omega}(I^\omega)$ relations measured with either increasing or decreasing $I^\omega$ are almost identical, and the $V^{3\omega}(I^\omega)$ relations are consistent within the small and large current range (see the inset in Figs. 2b, S5d, and Extended Data Fig. 4e).

(2) The third-order transport exhibits remarkable anisotropy that reflects the crystalline symmetry of (101)-$RuO_2$ thin films.

**Note 6: Calculation of nonlinear conductivity and nonlinearity**

Expanding the current density $J$ in response to an applied electric field $E$ (here $J$ and $E$ are both along a certain principal axis of a system) to the third order, we have

$$J = \sigma^{(1)}E + \sigma^{(2)}E^2 + \sigma^{(3)}E^3 \tag{S2}$$

$$I = \sigma^{(1)}Wt_{RO}\frac{V}{L} + \sigma^{(2)}Wt_{RO}(\frac{V}{L})^2 + \sigma^{(3)}Wt_{RO}(\frac{V}{L})^3 \tag{S3}$$

wherein $V$ and $I$ denote the amplitude of the applied voltage and the generated current, respectively; $\sigma^{(1)}$, $\sigma^{(2)}$, and $\sigma^{(3)}$ are the first-, second-, and third-order conductivity, respectively; $W$ and $L$ are the width of the current channel and the distance between two voltage probes of the Hall bar, respectively; $t_{RO}$ is the thickness of the RuO$_2$ layer. For simplicity, we rewrite Eq. S3 as

$$I = I^{(1)} + I^{(2)} + I^{(3)} = G^{(1)}V + G^{(2)}V^2 + G^{(3)}V^3 \tag{S4}$$

wherein $G^{(n)} = \sigma^{(n)}\frac{Wt_{RO}}{L^n}$ is the $n^{\text{th}}$-order conductance, $I^{(n)} = G^{(n)}V^n$ is the $n^{\text{th}}$-order current response, and $n = 1, 2, 3$. For the third-order response, we have

$$\sigma^{(3)} = \frac{I^{(3)}}{V^3}\frac{L^3}{Wt_{RO}} \tag{S5}$$

In this study, we detect the voltage in response to an applied current. Therefore, we need to deduce $I^{(3)}$ and $V$ to calculate $\sigma^{(3)}$.

Similar to Eqs. S2–S4, for the electric field generated by an applied current, we have

$$E = \rho^{(1)}J + \rho^{(2)}J^2 + \rho^{(3)}J^3 \tag{S6}$$

$$V = \rho^{(1)}L\frac{I}{Wt_{RO}} + \rho^{(2)}L(\frac{I}{Wt_{RO}})^2 + \rho^{(3)}L(\frac{I}{Wt_{RO}})^3 \tag{S7}$$

$$= V^{(1)} + V^{(2)} + V^{(3)} = R^{(1)}I + R^{(2)}I^2 + R^{(3)}I^3 \tag{S8}$$

Here $\rho^{(1)}$, $\rho^{(2)}$, and $\rho^{(3)}$ are the first-, second-, and third-order resistivity, respectively; $R^{(n)} = \frac{\rho^{(n)}L}{(Wt_{RO})^n}$ is the $n^{\text{th}}$-order resistance, $V^{(n)} = R^{(n)}I^n$ is the $n^{\text{th}}$-order voltage response, and $n = 1, 2, 3$. For an applied ac current $I = I_x^\omega \sin(\omega t)$ with amplitude $I_x^\omega$ and frequency $\omega$, the generated ac voltage is

$$V = R^{(1)}I_x^\omega \sin(\omega t) + R^{(2)}(I_x^\omega)^2 \sin^2(\omega t) + R^{(3)}(I_x^\omega)^3 \sin^3(\omega t)$$

$$= R^{(1)}I_x^\omega \sin(\omega t) + \frac{1}{2}R^{(2)}(I_x^\omega)^2[1 - \cos(2\omega t)] + \frac{1}{4}R^{(3)}(I_x^\omega)^3[3\sin(\omega t) - \sin(3\omega t)]$$

$$= \frac{1}{2}R^{(2)}(I_x^\omega)^2 + [R^{(1)}I_x^\omega + \frac{3}{4}R^{(3)}(I_x^\omega)^3]\sin(\omega t) + \frac{1}{2}R^{(2)}(I_x^\omega)^2\sin(2\omega t - \frac{\pi}{2}) - \frac{1}{4}R^{(3)}(I_x^\omega)^3\sin(3\omega t) \tag{S9}$$

Here we detect the amplitude of the third-harmonic voltage component with the lock-in technique.

$$V^{3\omega} = -\frac{1}{4}R^{(3)}(I_x^\omega)^3 = -\frac{\rho^{(3)}L}{4(Wt_{RO})^3}(I_x^\omega)^3 = -\frac{1}{4}V^{(3)} \tag{S10}$$

Note that $V^{3\omega}$ is proportional to the amplitude of $V^{(3)}$ with a factor of $-1/4$.

Since the second- and third-order responses are negligible compared to the first-order one, for the longitudinal response, we have

$$V \approx R^{(1)}I_x^\omega \tag{S11}$$

$$I^{(3)} \approx \frac{V^{(3)}}{R^{(1)}} \tag{S12}$$

Taking Eqs. S10–S12 into Eq. S5, the third-order longitudinal conductivity can be calculated as

$$\sigma_{//}^{(3)} = \frac{I^{(3)}}{V^3}\frac{L^3}{Wt_{RO}} = \frac{V_x^{3\omega}}{(I_x^\omega)^3}\frac{-4L^3}{Wt_{RO}[R^{(1)}]^4} \tag{S13}$$

For transverse response, Eq. S12 needs to be rewritten as

$$I^{(3)} \approx -\frac{V^{(3)}}{R^{(1)}}\frac{L}{W} \tag{S14}$$

The additional minus sign is due to our measurement geometry wherein a positive transverse voltage is induced by an electric field along the $+y$ direction. Taking Eqs. S10, S11, S14 into Eq. S5, the third-order Hall conductivity can be calculated as

$$\sigma_\perp^{(3)} = \frac{I^{(3)}}{V^3}\frac{L^3}{Wt_{RO}} = \frac{V_y^{3\omega}}{(I_x^\omega)^3}\frac{4L^4}{W^2 t_{RO}[R^{(1)}]^4} \tag{S15}$$

Similarly, the second-order Hall conductivity can be calculated as

$$\sigma_\perp^{(2)} = \frac{I^{(2)}}{V^2}\frac{L^2}{Wt_{RO}} = \frac{V_y^{2\omega}}{(I_x^\omega)^2}\frac{2L^3}{W^2 t_{RO}[R^{(1)}]^3} \tag{S16}$$

Note that in order to make a fair comparison with those reported in other studies, we omit the factors of 4 and 2 in Eqs. S13,15,16, respectively, when extracting the nonlinear conductivities in our work. Correspondingly, the first-principles calculation results (Eqs. 8–10 in the main text) include the constant factor as a denominator.

On the other hand, the nonlinearity can be calculated as

$$\frac{E_x^{3\omega}}{(E_x^\omega)^3} = \frac{V_x^{3\omega}}{(I_x^\omega)^3}\frac{L^2}{[R^{(1)}]^3} \tag{S17}$$

$$\frac{E_y^{3\omega}}{(E_x^\omega)^3} = \frac{V_y^{3\omega}}{(I_x^\omega)^3}\frac{L^3}{W[R^{(1)}]^3} \tag{S18}$$

$$\frac{E_y^{2\omega}}{(E_x^{\omega})^2} = \frac{V_y^{2\omega}}{(I_x^{\omega})^2}\frac{L^2}{W[R^{(1)}]^2} \tag{S19}$$

**Note 7: Determination of the third-order nonlinear Hall effect in thin samples**

Due to the unintentional misalignment of the Hall branches in a Hall cross, the as-measured $[V_y^{3\omega}]_{\text{raw}}$ contains contributions from the strong third-order longitudinal transport effect. For ultrathin samples with a large first-order resistance but small in-plane resistivity anisotropy, such contribution is non-negligible because the transverse resistance is mainly induced by the misalignment. In order to obtain a clean third-order nonlinear Hall effect, we first deduced the misalignment-induced first-order resistance $R_x^{(1)}$ of the Hall cross and the four-probe resistance $R_y^{(1)}$ of the sample from first-order harmonic Hall and longitudinal transport data, respectively. Note that according to Eq. S9, the collected $V^\omega$ contains contributions from both the first- and the third-order resistance, and the latter is larger than the $V^{3\omega}$ with a factor of –3. Therefore, we fit the $V^\omega(I_x^\omega)$ relations with the following equation.

$$V^\omega = R^{(1)} I_x^\omega + \frac{3}{4} R^{(3)} (I_x^\omega)^3 \tag{S20}$$

As displayed in Fig. S3a, the $R_x^{(1)}$ and $R_y^{(1)}$ for a typical sample can be deduced from the fitting. Moreover, the contribution from $R_x^{(3)}$ to $V_x^\omega$ is indeed ~–3 times larger than the $V_x^{3\omega}$ (Fig. S3b).

Note that the in-(101)-plane resistivity anisotropy of RuO$_2$ is subtle, and assuming that the misaligned Hall branches are separated by a distance $\delta L$, we have

$$\frac{R_y^{(1)}}{R_x^{(1)}} \approx \frac{\delta L}{L} \tag{S21}$$

On the other hand, according to Eq. S7, the contribution to $[V_y^{3\omega}]_{\text{raw}}$ from $V_x^{3\omega}$ due to misalignment is $\frac{\delta L}{L} V_x^{3\omega}$. Therefore, based on Eq. S21, the genuine $V_y^{3\omega}$ can be calculated as

$$V_y^{3\omega} = [V_y^{3\omega}]_{\text{raw}} - \frac{R_y^{(1)}}{R_x^{(1)}} V_x^{3\omega} \tag{S22}$$

## Note 8: Crystalline anisotropy of third-order nonlinearity

The symmetry elements of a rutile (101)-RuO$_2$ slab comprise a glide mirror plane $\widetilde{\mathcal{M}}_{010}$ and an inversion center, which is similar to the case of a few-layer WTe$_2$ (ref. [42]). As only in-film-plane electrical transport properties can be investigated, the first-order electric field generated by an in-plane current $\boldsymbol{J} = J\begin{pmatrix}\cos\Psi\\ \sin\Psi\end{pmatrix}$ can be expressed with a two-dimensional resistivity tensor $\boldsymbol{\rho}^{(1)} = \begin{pmatrix}\rho_b & 0\\ 0 & \rho_a\end{pmatrix}$ as $\boldsymbol{E}^{(1)} = \boldsymbol{\rho}^{(1)}\boldsymbol{J} = J\begin{pmatrix}\rho_b\cos\Psi\\ \rho_a\sin\Psi\end{pmatrix}$, wherein $\Psi$ is the angle between $\boldsymbol{J}$ and [010]; $\rho_b$ and $\rho_a$ are the longitudinal resistivities of RuO$_2$ along [010] and [$\bar{1}$01], respectively. As a result, the first-order longitudinal electric field $E_{//}^{(1)}$ generated by $\boldsymbol{J}$ is

$$E_{//}^{(1)} = J(\rho_b\cos^2\Psi + \rho_a\sin^2\Psi) \tag{S23}$$

Correspondingly, the first-order longitudinal resistivity of (101)-RuO$_2$ as a function of $\Psi$ is

$$\rho_{//}^{(1)} = \rho_b\cos^2\Psi + \rho_a\sin^2\Psi \tag{S24}$$

On the other hand, for third-order nonlinear transport effects, the third-order current density $J_i^{(3)}$ in response to applied electric fields is $J_i^{(3)} = \sigma_{ijkl}E_jE_kE_l$, wherein $\sigma_{ijkl}$ is the corresponding component of the third-order conductivity tensor with $i$–$j$ denoting the [010] ($b$) or [$\bar{1}$01] ($a$) axis. As a result of symmetry constraints, there are only four independent $\sigma_{ijkl}$ for a (101)-RuO$_2$ thin film, *i.e.*, $\sigma_1^{(3)} = \sigma_{bbbb}$, $\sigma_2^{(3)} = \sigma_{aaaa}$, $\sigma_3^{(3)} = (\sigma_{bbaa} + \sigma_{baba} + \sigma_{baab})/3$, and $\sigma_4^{(3)} = (\sigma_{aabb} + \sigma_{abab} + \sigma_{abba})/3$. Therefore, the third-order electric field $\boldsymbol{E}^{(3)}$ generated by $\boldsymbol{J}$ is $\boldsymbol{E}^{(3)} \approx \boldsymbol{\rho}^{(1)}\boldsymbol{J}^{(3)} = J^3\begin{pmatrix}\sigma_1^{(3)}\rho_b^4\cos^3\Psi + 3\sigma_3^{(3)}\rho_b^2\rho_a^2\cos\Psi\sin^2\Psi\\ \sigma_2^{(3)}\rho_a^4\sin^3\Psi + 3\sigma_4^{(3)}\rho_b^2\rho_a^2\sin\Psi\cos^2\Psi\end{pmatrix}$.

The third-order nonlinearity as a function of $\Psi$ can be expressed as

$$\frac{E_x^{3\omega}}{(E_x^{\omega})^3} = \frac{\sigma_1^{(3)}\rho_b^4\cos^4\Psi + \sigma_2^{(3)}\rho_a^4\sin^4\Psi + 3[\sigma_3^{(3)} + \sigma_4^{(3)}]\rho_b^2\rho_a^2\sin^2\Psi\cos^2\Psi}{(\rho_b\cos^2\Psi + \rho_a\sin^2\Psi)^3} \tag{S25}$$

$$\frac{E_y^{3\omega}}{(E_x^{\omega})^3} = \frac{[3\sigma_4^{(3)}\rho_b^2\rho_a^2 - \sigma_1^{(3)}\rho_b^4]\sin\Psi\cos^3\Psi + [\sigma_2^{(3)}\rho_a^4 - 3\sigma_3^{(3)}\rho_b^2\rho_a^2]\sin^3\Psi\cos\Psi}{(\rho_b\cos^2\Psi + \rho_a\sin^2\Psi)^3} \tag{S26}$$

Note that $E_x^{3\omega}/(E_x^{\omega})^3 = \rho_b\sigma_1^{(3)}$ at $\Psi = 0°$ while $E_x^{3\omega}/(E_x^{\omega})^3 = \rho_a\sigma_2^{(3)}$ at $\Psi = 90°$. Utilizing these constraints, the crystalline anisotropy of the first-order longitudinal resistance and the third-order nonlinearity can be fitted with Eqs. S24–26.

**Note 9: Disorder scattering effects on third-order electrical transport**

***Boltzmann equation in the third-order regime***: The nonlinear electron current density is expressed as

$$\boldsymbol{j} = -e \sum_l f_l \boldsymbol{v}_l \tag{S27}$$

where $l \equiv (n, \boldsymbol{k})$ is a combination of Bloch band index $n$ and momentum $\boldsymbol{k}$. When considering the scattering effect, side-jump effect could cause a positional shift. The velocity $\boldsymbol{v}_l$ is obtained as

$$\boldsymbol{v}_l = \frac{\partial}{\hbar \partial_{\boldsymbol{k}}} \tilde{\varepsilon}_l + \boldsymbol{E} \times \tilde{\boldsymbol{\Omega}}_l + \boldsymbol{v}_l^{sj} \tag{S28}$$

where $\tilde{\boldsymbol{\Omega}}_l = \boldsymbol{\Omega}_l + \boldsymbol{\Omega}_l^{(1)} + \boldsymbol{\Omega}_l^{(2)}$ is the Berry curvature expanded to $E^2$, and $\boldsymbol{v}_l^{sj}$ is the side-jump velocity. The distribution function $f_l$ under electric field $\boldsymbol{E}$ at static limit is determined with the semiclassical Boltzmann equation

$$-\boldsymbol{E} \cdot \partial_{\boldsymbol{k}} f_l = \hat{I}_{coll} f_l \tag{S29}$$

where $\hat{I}_{coll}$ is the collision integral for scattering. In the perturbative treatment of electric field, $f_l$ can be obtained with ascending powers of $E$. Define $f_l^{(p,-q)}$ as the term in $f_l$ of the order $E^p V^{-q}$.

$$f_l = f_l^0 + \sum_{p=1,2,3;q} f_l^{(p,-q)} \tag{S30}$$

where $f_l^0$ is the Fermi-Dirac distribution and $V$ is the disorder strength. Following the same procedure for $f_l$ expanded to the $E^2$ (refs. [40,41,43]), we derive the new equations for $f_l$ expanded at $E^3$. For $f_l^{(3,-q)}, (q = 0,1,...,2p)$, the equation series read

$$-D_E f_l^{(2,-4)} = \sum_{l'} \varpi_{l'l}^s \left[ f_l^{(3,-6)} - f_{l'}^{(3,-6)} \right] \tag{S31}$$

$$-D_E f_l^{(2,-3)} = \sum_{l'} \varpi_{l'l}^s \left[ f_l^{(3,-5)} - f_{l'}^{(3,-5)} \right] + \sum_{l'} \varpi_{l'l}^{(3)as} \left[ f_l^{(3,-6)} + f_{l'}^{(3,-6)} \right] \tag{S32}$$

$$-D_E f_l^{(2,-2)} = \sum_{l'} \delta^E \varpi_{l'l}^s \left[ f_l^{(2,-4)} - f_{l'}^{(2,-4)} \right] + \sum_{l'} \varpi_{l'l}^s \left[ f_l^{(3,-4)} - f_{l'}^{(3,-4)} \right] +$$
$$\sum_{l'} \varpi_{l'l}^{(3)as} \left[ f_l^{(3,-5)} + f_{l'}^{(3,-5)} \right] + \sum_{l'} \varpi_{l'l}^{(4)as} \left[ f_l^{(3,-6)} + f_{l'}^{(3,-6)} \right] \tag{S33}$$

$$-D_E f_l^{(2,-1)} = \sum_{l'} \delta^E \varpi_{l'l}^s \left[ f_l^{(2,-3)} - f_{l'}^{(2,-3)} \right] + \sum_{l'} \varpi_{l'l}^s \left[ f_l^{(3,-3)} - f_{l'}^{(3,-3)} \right] + \sum_{l'} \delta^E \varpi_{l'l}^{(3)as} \left[ f_l^{(2,-4)} + \right.$$

$$f_{l'}^{(2,-4)}\Big] + \sum_{l'} \varpi_{l'l}^{(3)as}\left[f_l^{(3,-4)} + f_{l'}^{(3,-4)}\right] + \sum_{l'} \varpi_{l'l}^{(4)as}\left[f_l^{(3,-5)} + f_{l'}^{(3,-5)}\right] \tag{S34}$$

$$-D_E f_l^{(2,0)} = \sum_{l'} \delta^{E^2} \varpi_{l'l}^{S}\left[f_l^{(1,-2)} - f_{l'}^{(1,-2)}\right] + \sum_{l'} \delta^{E} \varpi_{l'l}^{S}\left[f_l^{(2,-2)} - f_{l'}^{(2,-2)}\right] +$$

$$\sum_{l'} \varpi_{l'l}^{S}\left[f_l^{(3,-2)} - f_{l'}^{(3,-2)}\right] + \sum_{l'} \delta^{E} \varpi_{l'l}^{(3)as}\left[f_l^{(2,-3)} + f_{l'}^{(2,-3)}\right] + \sum_{l'} \varpi_{l'l}^{(3)as}\left[f_l^{(3,-3)} + f_{l'}^{(3,-3)}\right] +$$

$$\sum_{l'} \delta^{E} \varpi_{l'l}^{(4)as}\left[f_l^{(2,-4)} + f_{l'}^{(2,-4)}\right] + \sum_{l'} \varpi_{l'l}^{(4)as}\left[f_l^{(3,-4)} + f_{l'}^{(3,-4)}\right] \tag{S35}$$

$$-D_E f_l^{(2,1)} = \sum_{l'} \delta^{E^2} \varpi_{l'l}^{S}\left[f_l^{(1,-1)} - f_{l'}^{(1,-1)}\right] + \sum_{l'} \delta^{E} \varpi_{l'l}^{S}\left[f_l^{(2,-1)} - f_{l'}^{(2,-1)}\right] +$$

$$\sum_{l'} \varpi_{l'l}^{S}\left[f_l^{(3,-1)} - f_{l'}^{(3,-1)}\right] + \sum_{l'} \delta^{E^2} \varpi_{l'l}^{(3)as}\left[f_l^{(1,-2)} + f_{l'}^{(1,-2)}\right] + \sum_{l'} \delta^{E} \varpi_{l'l}^{(3)as}\left[f_l^{(2,-2)} +$$

$$f_{l'}^{(2,-2)}\right] + \sum_{l'} \varpi_{l'l}^{(3)as}\left[f_l^{(3,-2)} + f_{l'}^{(3,-2)}\right] + \sum_{l'} \delta^{E} \varpi_{l'l}^{(4)as}\left[f_l^{(2,-3)} + f_{l'}^{(2,-3)}\right] +$$

$$\sum_{l'} \varpi_{l'l}^{(4)as}\left[f_l^{(3,-3)} + f_{l'}^{(3,-3)}\right] \tag{S36}$$

$$-D_E f_l^{(2,2)} = \sum_{l'} \delta^{E^3} \varpi_{l'l}^{S}[f_l^0 - f_{l'}^0] + \sum_{l'} \delta^{E^2} \varpi_{l'l}^{S}\left[f_l^{(1,0)} - f_{l'}^{(1,0)}\right] + \sum_{l'} \delta^{E} \varpi_{l'l}^{S}\left[f_l^{(2,0)} -\right.$$

$$\left.f_{l'}^{(2,0)}\right] + \sum_{l'} \varpi_{l'l}^{S}\left[f_l^{(3,0)} - f_{l'}^{(3,0)}\right] + \sum_{l'} \delta^{E^2} \varpi_{l'l}^{(3)as}\left[f_l^{(1,-1)} + f_{l'}^{(1,-1)}\right] +$$

$$\sum_{l'} \delta^{E} \varpi_{l'l}^{(3)as}\left[f_l^{(2,-1)} + f_{l'}^{(2,-1)}\right] + \sum_{l'} \varpi_{l'l}^{(3)as}\left[f_l^{(3,-1)} + f_{l'}^{(3,-1)}\right] + \sum_{l'} \delta^{E^2} \varpi_{l'l}^{(4)as}\left[f_l^{(1,-2)} +\right.$$

$$\left.f_{l'}^{(1,-2)}\right] + \sum_{l'} \delta^{E} \varpi_{l'l}^{(4)as}\left[f_l^{(2,-2)} + f_{l'}^{(2,-2)}\right] + \sum_{l'} \varpi_{l'l}^{(4)as}\left[f_l^{(3,-2)} + f_{l'}^{(3,-2)}\right] \tag{S37}$$

where $D_E \equiv \boldsymbol{E} \cdot \partial_{\boldsymbol{k}}$. The scattering rate parameters $\varpi_{l'l}^{S}$ are the symmetric part of $\varpi_{l'l}^{(2)} = \frac{2\pi}{\hbar}\langle|V_{ll'}|^2\rangle_{dis}\delta(\varepsilon_l - \varepsilon_{l'})$, where $V_{ll'}$ is the scattering matrix element and $\langle\cdots\rangle_{dis}$ is the average over disorders. $\delta^{E^n}\varpi_{l'l}^{S}$ is its nth-order electric-field correction. $\varpi_{l'l}^{(3)as}$ and $\varpi_{l'l}^{(4)as}$ are the antisymmetric part of $\varpi_{l'l}^{(3)}$ and $\varpi_{l'l}^{(4)}$ (see ref. [40] for details), respectively. The terms including $\delta^{E^n}\varpi_{l'l}^{S}$ and $\boldsymbol{v}_l^{sj}$ contribute to the side-jump effect, while the terms including $\varpi_{l'l}^{as} = \varpi_{l'l}^{(3)as} + \varpi_{l'l}^{(4)as}$ and their electric field corrections contribute to the skew-scattering effect.

***$\mathcal{T}$-odd contributions from scattering effects***: A full solution of $f_l^{(p,-q)}$ and the third-order conductivity is tedious and not the focus of our work. The scattering effect in time-reversal invariant ($\mathcal{T}$-even) systems has been studied in ref. [44]. Here we propose several $\mathcal{T}$-odd contributions from high-order skew-scattering and side-jump effects to third-order conductivity. The first- and second-order skew-scattering effects at static limit are expressed as

$$\sigma_{abcd}^{SK} = \frac{e^4\tau^2}{\hbar^2}\sum_{ll'}\left\{(\partial_a G_{bc,l} - \partial_b G_{ac,l} - \partial_a G_{bc,l'} + \partial_b G_{ac,l'})\varpi_{l'l}^{(4)as}\partial_c f_l^0\right\} \qquad (S38)$$

$$\sigma_{abcd}^{2SK} = \frac{e^4\tau^4}{\hbar^2}\sum_{ll'l''}\epsilon^{abe}\left\{(\partial_c\Omega_l^e - \partial_c\Omega_{l'}^e)\varpi_{l'l}^{(4)as}\varpi_{l''l'}^{(4)as} + \Omega_e(\partial_c - \partial_{c'})\varpi_{l'l}^{(4)as}\varpi_{l''l'}^{(4)as}\right\}\partial_d f_l^0 \quad (S39)$$

and the second-order side-jump effect is

$$\sigma_{abcd}^{2SJ} = \frac{e^3\tau^2}{\hbar}\sum_{l,l'}\epsilon^{abe}\left(\Omega_l^e - \Omega_{l'}^e\right)[\Lambda_{ll',c}\delta(\varepsilon_l - \varepsilon_{l'}) + O_{ll',c}]v_{l,d}^{sj}\partial_{\varepsilon_l}f_0 \qquad (S40)$$

where $\partial_a \equiv \frac{\partial}{\partial k_a}$, $\Lambda_{ll',a} = 4\pi Re(\sum_{m\neq l'}\frac{\langle u_m|i\partial_a u_{l'}\rangle\langle V_{l'l}V_{lm}\rangle_{dis}}{\varepsilon_{l'}-\varepsilon_m} + \sum_{m\neq l}\frac{\langle V_{ll'}V_{l'm}\rangle_{dis}\langle u_m|i\partial_a u_l\rangle}{\varepsilon_l-\varepsilon_m})$ and $O_{ll',a} = 2\pi\langle|V_{ll'}|^2\rangle_{dis}\frac{\partial\delta(\varepsilon_l-\varepsilon_{l'})}{\partial\varepsilon_l}\delta r_{ll',a}$. $\Omega_l^a$ and $\frac{1}{2}G_{ab,l}$ are the Berry curvature and band resolved quantum metric tensor for $l$ state, respectively. The above effects contribute solely to the Hall current.

From Matthiessen's rule, when there are multiple types of scattering sources, $\rho_{xx} = \sum_i \rho_i$, where $\rho_i$ is the partial resistivity. The scaling of the above contributions with partial resistivities can be expressed as $\sigma_{abcd}^{SK} \sim \sum_{ij} C_{ij}^{SK}\rho_i\rho_j/\rho_{xx}^2$, $\sigma_{abcd}^{2SK} \sim \sum_{ijkl} C_{ijkl}^{2SK}\rho_i\rho_j\rho_k\rho_l/\rho_{xx}^4$ and $\sigma_{abcd}^{2SJ} \sim \sum_{ij} C_{ij}^{2SJ}\rho_i\rho_j/\rho_{xx}^2$.

**Note 10: Third-order electrical transport in thick samples**

Since the altermagnetism in RuO$_2$ is likely absent in bulk, we here try to explore a possible thickness-driven altermagnetic-paramagnetic transition via monitoring the third-order nonlinear transport responses. Note that defect density and, possibly, magnetic properties can change with the thickness of RuO$_2$ thin films, both of which would affect the nonlinear transport. As shown in Fig. S10a, the room-temperature conductivity of RuO$_2$ thin films is less dependent on thickness for samples thicker than 20 nm. This suggests that the defect density could be assumed almost constant within this range, so that the variation in the nonlinear transport properties might be ascribed to the possible magnetic transition. Given that the 20-nm-thick sample is likely altermagnetic due to the existence of anisotropic exchange bias (Fig. 2c in [*Adv. Mater.* **37**, 2507764 (2025)]), we compare the nonlinear transport properties in this sample and in an 80-nm-thick layer. Considering that $\mathcal{T}$-odd effects are characteristic of altermagnetic order, we mainly focused on the third-order transverse responses.

Figure S10b depicts the third-harmonic voltage $V_y^{3\omega}$ generated by an ac current applied along $[\bar{1}11]^*$ in the two samples. Notably, the sign of $V_y^{3\omega}$ is opposite to the 8-nm-thick samples in the main text, which can stem from distinct domain population/distribution, or simply from attenuated $\mathcal{T}$-odd contributions. We then perform scaling analyses at 200–300 K to shed light on the microscopic mechanisms. Since the longitudinal conductivity $\sigma_{xx}$ of the two samples are comparable in the studied temperature range (Fig. S10c), the scaling results can be directly compared. As displayed in Fig. S10d, the third-order transverse conductivity $\sigma_\perp^{(3)}$ scales linearly with the first-order longitudinal conductivity $\sigma_{xx}$ in the 20-nm-thick film, similar to the case of the 8-nm-thick samples in the main text. In contrast, the $\sigma_\perp^{(3)}$ depends on $\sigma_{xx}$ in a nonlinear manner in the 80-nm-thick layer, suggestive of changed origins of nonlinear transverse responses due to the possible magnetic transition.

The $\sigma_\perp^{(3)}(\sigma_{xx})$ relation of the 20-nm-thick film can be well fitted with

$$\sigma_\perp^{(3)} = \zeta \sigma_{xx} + \lambda \tag{S41}$$

wherein $\zeta$ and $\lambda$ are fitting constants. Note that the $\lambda$ term reflects the contribution to $\sigma_\perp^{(3)}$ from $\mathcal{T}$- odd mechanisms. On the other hand, the $\sigma_\perp^{(3)}(\sigma_{xx})$ relation showing certain degree of nonlinearity of the 80-nm-thick film is fitted with

$$\sigma_\perp^{(3)} = \gamma\sigma_{xx}^3 + \zeta\sigma_{xx} + \lambda \tag{S42}$$

wherein $\gamma$ is a fitting constant and the $\gamma\sigma_{xx}^3$ term stands for additional $\mathcal{T}$-even contributions to $\sigma_\perp^{(3)}$ from the paramagnetic bulk state. According to the fitting results, the $\mathcal{T}$-odd component of $|\sigma_\perp^{(3)}|$ decreases from ~25.1 μm A V$^{-3}$ in the 20-nm-thick layer to ~14.2 μm A V$^{-3}$ in the 80-nm-thick one, indicating weakened $\mathcal{T}$-odd mechanisms with increasing thickness. Moreover, the $\sigma_\perp^{(3)}(\sigma_{xx})$ relation for the thick sample can even be fitted with

$$\sigma_\perp^{(3)} = \gamma\sigma_{xx}^3 + \zeta\sigma_{xx} \tag{S43}$$

wherein the $\mathcal{T}$-odd contribution is completely omitted. In correspondence with the attenuated $\mathcal{T}$-odd components of $\sigma_\perp^{(3)}$, the third-order transverse responses collected with different Hall crossings in the same 80-nm-thick device are almost identical (Fig. S10e), suggesting the possible degradation of altermagnetic domains. In particular, the nonlinearity $E_y^{3\omega}/(E_x^\omega)^3$ only changes by a factor of ~1.06, much smaller than that of ~1.50 for the 8-nm-thick sample (derived from Fig. 2g in the main text). The above features suggest, although inconclusively, the attenuation of altermagnetism in the 80-nm-thick film.

**Supplementary Figures**

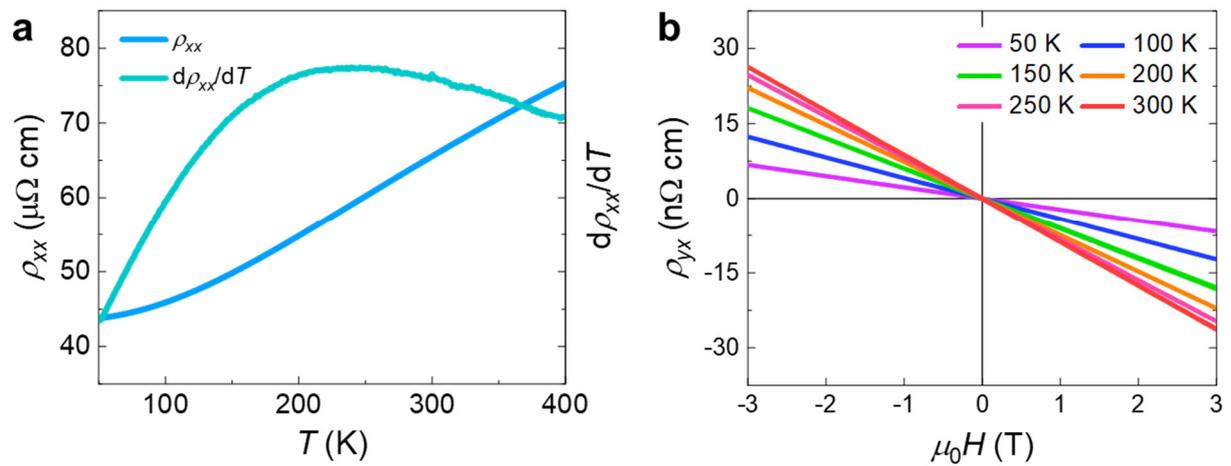

**Fig. S1 | Basic dc transport properties of (101)-oriented RuO$_2$ thin films. a**, Temperature $T$ dependence of the longitudinal resistivity $\rho_{xx}$ and its first derivative (d$\rho_{xx}$/d$T$) measured with a typical Hall bar device based on an 8-nm-thick (101)-RuO$_2$ thin film. **b**, Magnetic-field $\mu_0 H$ dependence of Hall resistivity $\rho_{yx}$ at various $T$ of the sample.

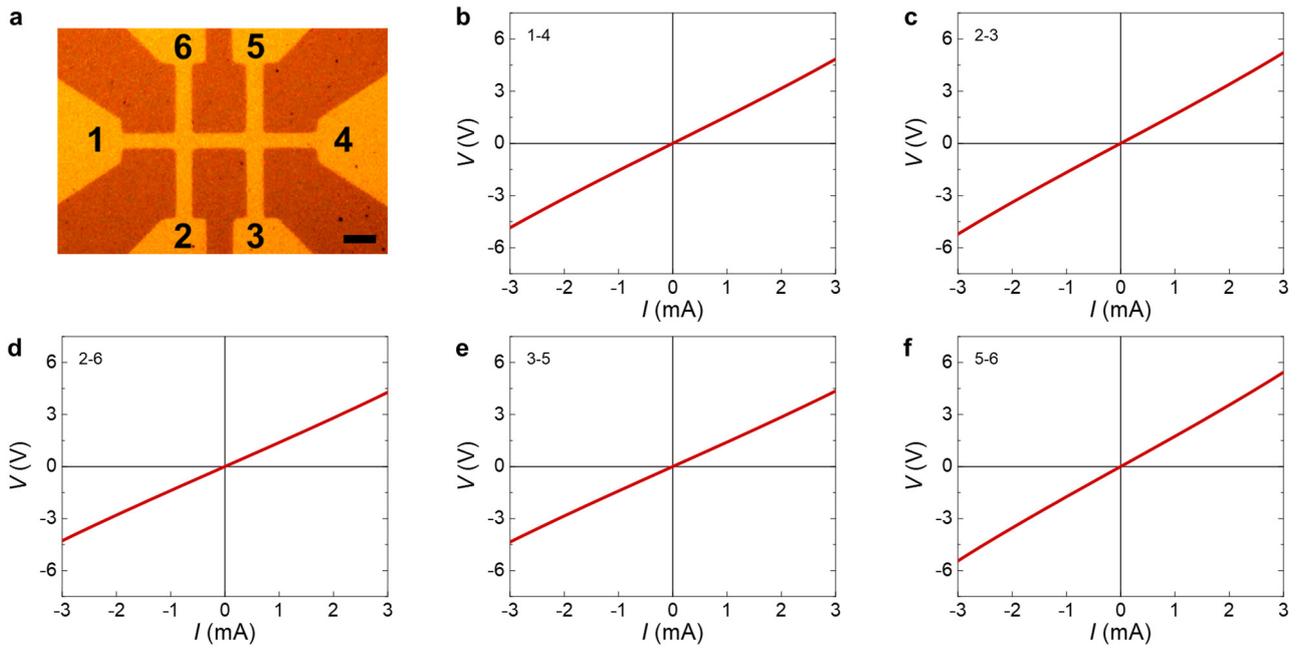

**Fig. S2 | Two-probe dc transport property of a typical device based on an 8-nm-thick (101)-oriented RuO$_2$ thin film. a**, Optical image of a typical Hall bar device. The scale bar represents a length of 5 μm. **b**–**f**, Two-probe dc voltage $V$ as a function of applied current $I$ measured from different pair of electrodes.
25

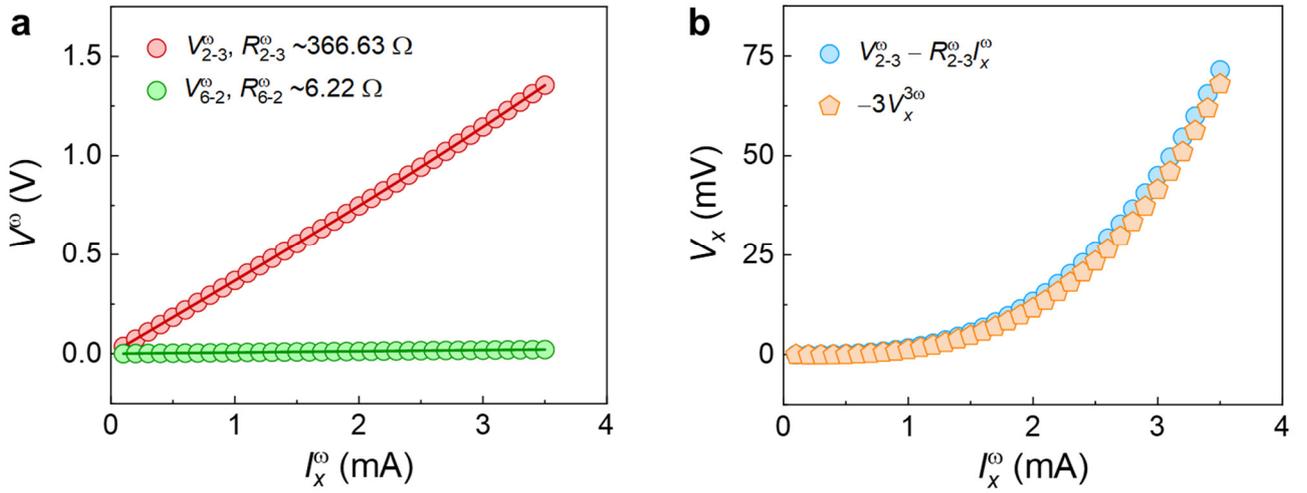

**Fig. S3 | Determination of the first-order resistance. a**, First-harmonic voltage $V^\omega$ as a function of $I_x^\omega$ measured from a pair of longitudinal and Hall electrodes of a typical Hall bar device based on an 8-nm-thick (101)-oriented RuO$_2$ thin film. The solid lines are fittings to the data according to Eq. S20. **b**, Relation between the third-order-resistance contribution to the first-harmonic longitudinal voltage $V_x^\omega$ and the third-harmonic longitudinal voltage $V_x^{3\omega}$.

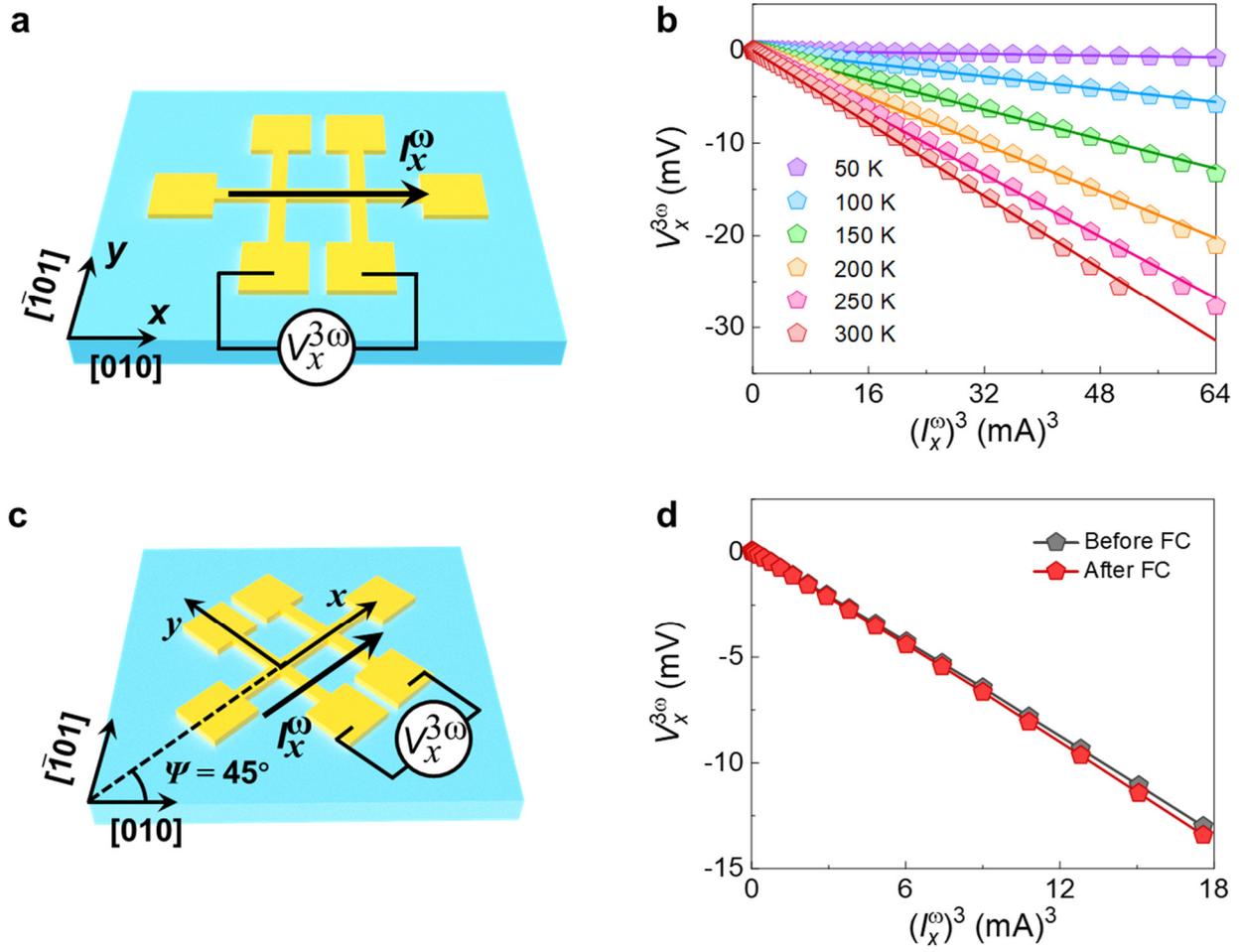

**Fig. S4 | Additional data for the third-order longitudinal electrical transport in an 8-nm-thick (101)-oriented RuO$_2$ thin film. a** and **c**, Schematic illustration on the measurement protocol of the third-order longitudinal transport. **b**, Third-harmonic longitudinal voltages $V_x^{3\omega}$ as a function of $(I_x^{\omega})^3$ at various temperature. $I_x^{\omega}$ is applied along [010], corresponding to **a**. **d**, $V_x^{3\omega}$ as a function of $(I_x^{\omega})^3$ measured before and after field cooling (FC) the sample with a magnetic field of 30 T from 423 K. $I_x^{\omega}$ is applied along $[\bar{1}11]^*$, corresponding to **c**. The solid lines in **b** are linear fittings to the data.

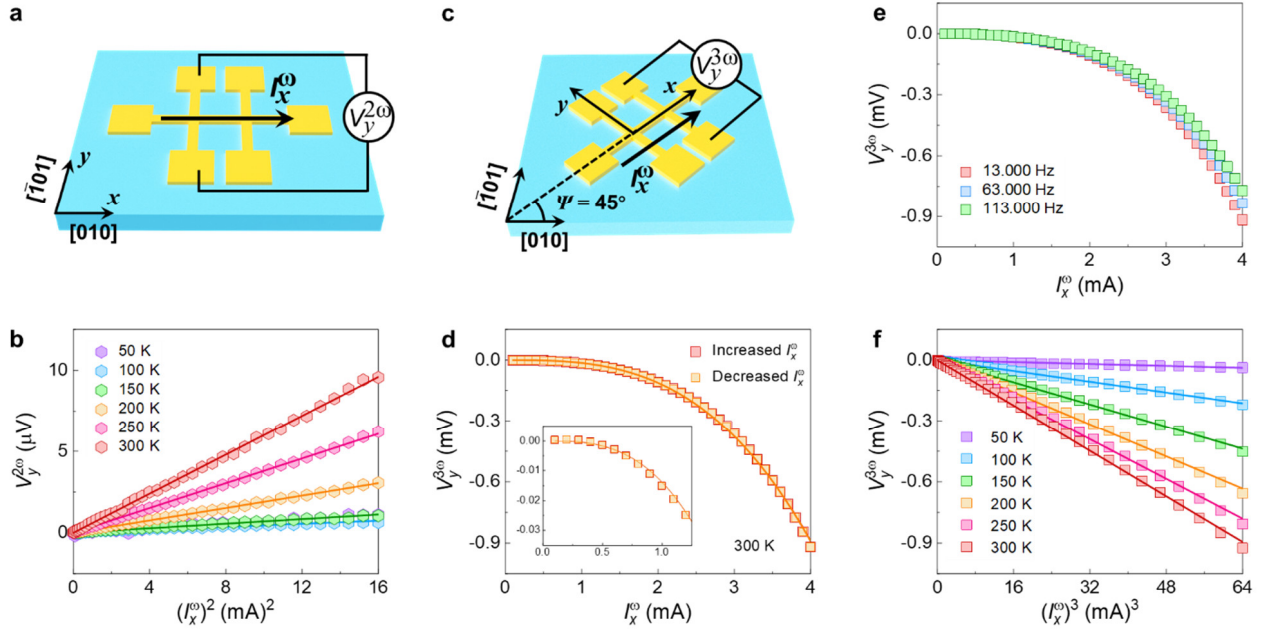

**Fig. S5 | Additional data for the second- and third-order transverse electrical transport in an 8-nm-thick (101)-oriented RuO$_2$ thin film. a**, Schematic illustration on the measurement protocol of the second-order Hall effect. **b**, Second-harmonic transverse voltages $V_y^{2\omega}$ as a function of $(I_x^\omega)^2$ recorded at various temperatures. $I_x^\omega$ is applied along [010]. **c**, Schematic illustration on the measurement protocol of the third-order Hall effect. **d–f**, Third-harmonic transverse voltages $V_y^{3\omega}$ as a function of $I_x^\omega$ or $(I_x^\omega)^3$ measured in distinct conditions. $I_x^\omega$ is applied along $[\bar{1}11]^*$. The inset in **d** displays the data collected with a small $I_x^\omega$. The solid lines in **b** and **f** are linear fittings to the data. The solid lines in **d** are cubic fittings to the data.

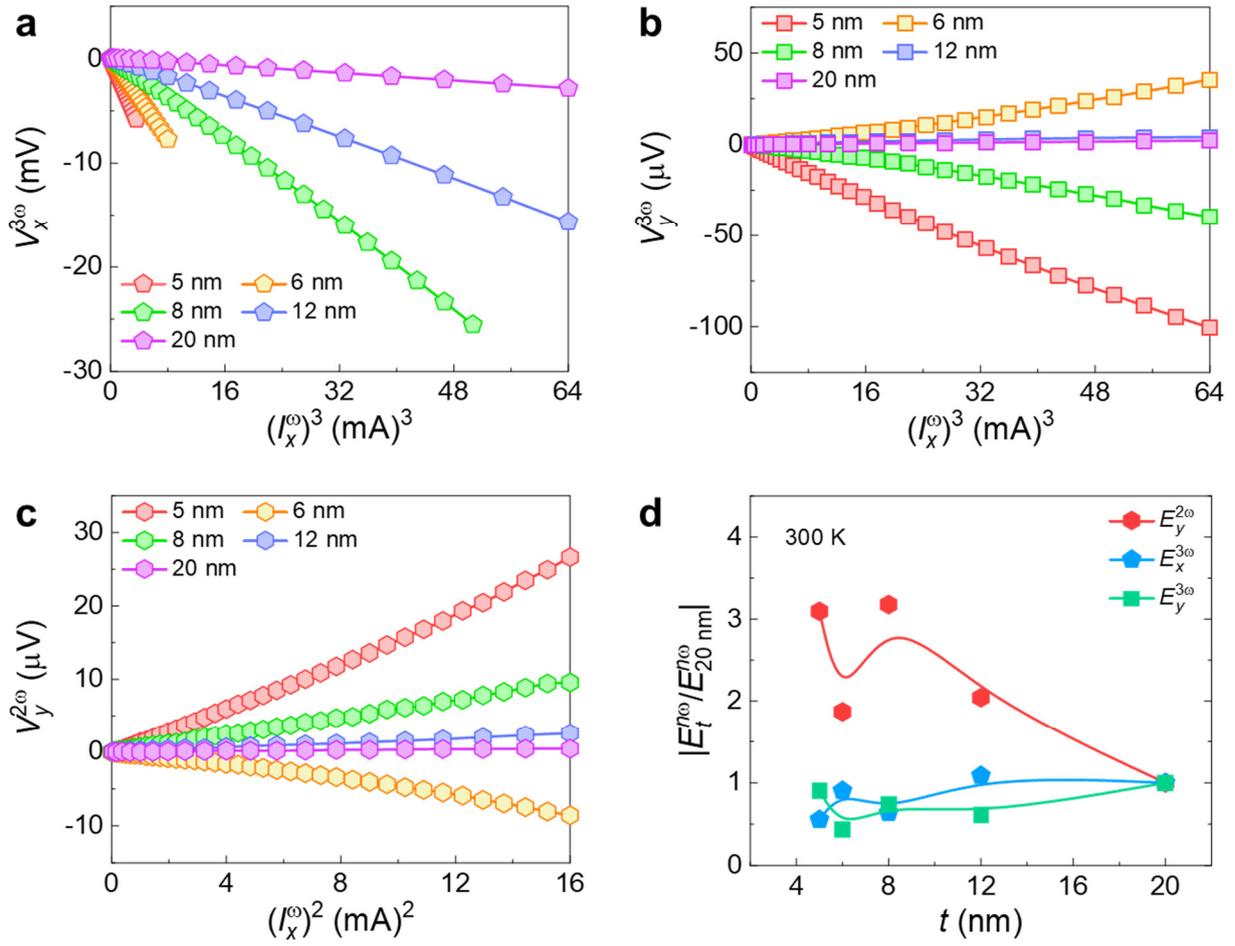

**Fig. S6 | Thickness dependence of the nonlinear electrical transport in (101)-oriented RuO$_2$ thin films. a–c**, Third-harmonic longitudinal voltage $V_x^{3\omega}$, $V_y^{3\omega}$, and $V_y^{2\omega}$ as a function of $(I_x^\omega)^3$ or $(I_x^\omega)^2$ in samples with thickness ranging from 5 to 20 nm at room temperature. $I_x^\omega$ is applied along [010]. Owing to the voltage input limit of the lock-in amplifier, the maximum $I_x^\omega$ in the $V_x^{3\omega}$ measurement is changed for samples with different thickness. **d**, Thickness-dependent second- and third-harmonic electric fields $E^{2\omega}$ and $E^{3\omega}$ generated by $I_x^\omega$ with a fixed current density. The data are extracted from **a**–**c** and normalized by the results of the 20-nm-thick sample.

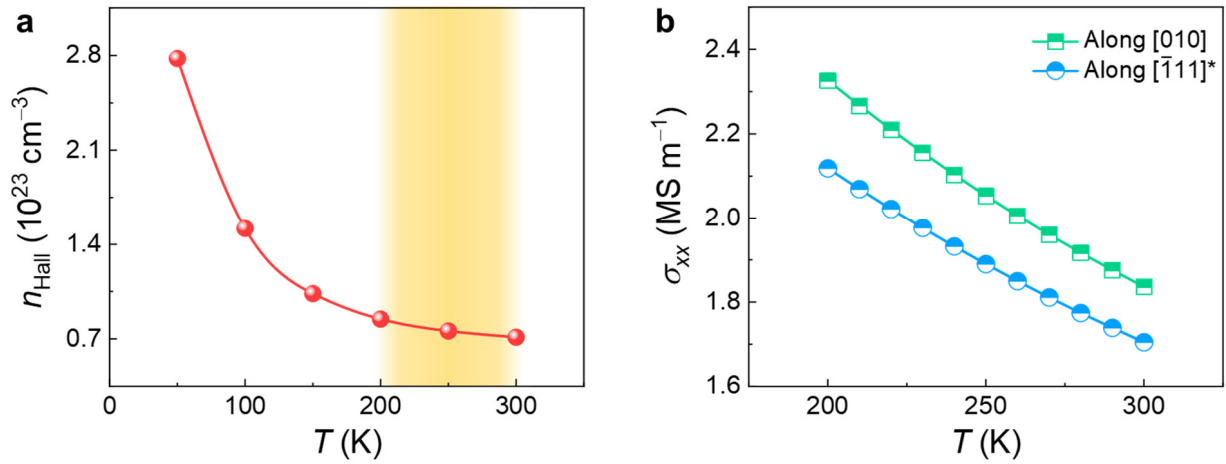

**Fig. S7 | Transport properties of typical devices based on 8-nm-thick (101)-oriented RuO$_2$ thin films. a**, Hall carrier density $n_{Hall}$ deduced from Fig. S1**b** as a function of $T$. **b**, Longitudinal conductivity $\sigma_{xx}$ along [010] and $[\bar{1}11]^*$ (collected with two different samples) extracted from first-harmonic measurements at 200–300 K.

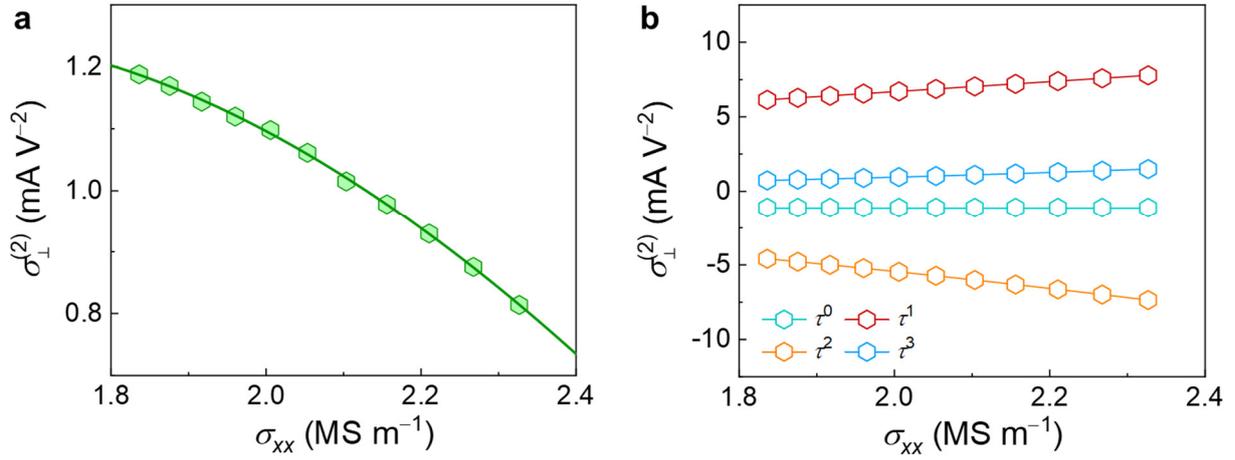

**Fig. S8 | Scaling relation of the second-order Hall effect in an 8-nm-thick (101)-oriented RuO$_2$ thin film. a**, Relation between the second-order Hall conductivity $\sigma_\perp^{(2)}$ and $\sigma_{xx}$ measured with $I_x^\omega$ along [010] at 200–300 K. The solid line is fitting to the data according to Eq. S1. **b**, Contributions to $\sigma_\perp^{(2)}$ from different mechanisms that exhibit distinct dependence on electron relaxation time $\tau$. The data are derived according to the fitting in **a**.

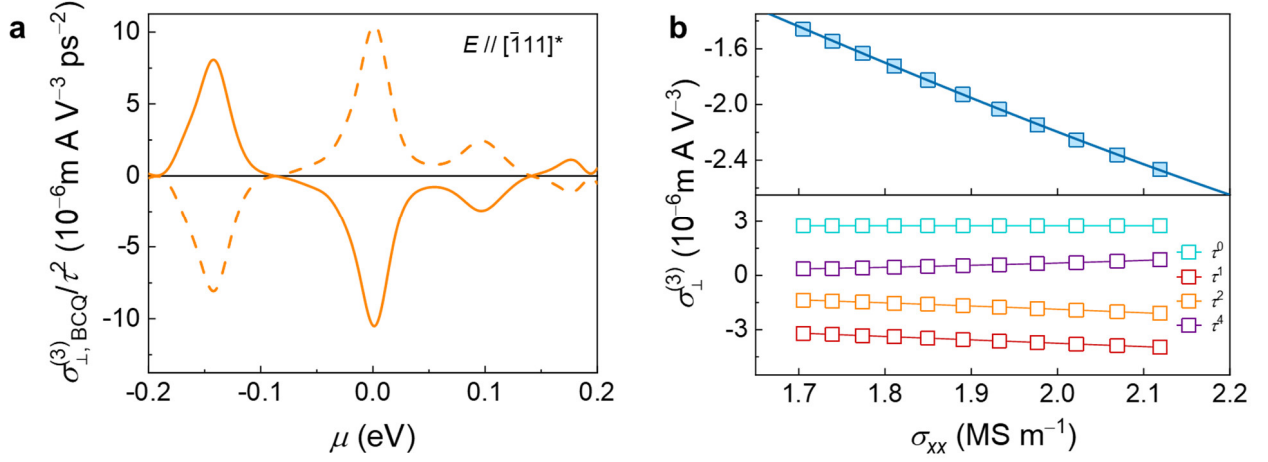

**Fig. S9 | Scaling relation of the third-order Hall effect in an 8-nm-thick (101)-oriented RuO$_2$ thin film incorporating multiple $\mathcal{T}$-odd mechanisms. a**, First-principles calculation results of third-order Hall conductivity contributed by Berry curvature quadrupoles $\sigma^{(3)}_{\perp,\text{BCQ}}$ divided by $\tau^2$. $\tau$ with a typical value of ~0.04 ps denotes electron scattering time. The dash line denotes the results after time reversal. $I^\omega_x$ is applied along $[\bar{1}11]^*$ in the calculation. **b**, Experimentally-obtained third-order Hall conductivity $\sigma^{(3)}_\perp$ versus $\sigma_{xx}$ at 200–300 K (upper panel). The solid line is fitting to the data. The lower panel display the respective contributions to $\sigma^{(3)}_\perp$ that are proportional to $\tau^0$, $\tau^1$, $\tau^2$, and $\tau^4$. $I^\omega_x$ is applied along $[\bar{1}11]^*$ in the measurement.

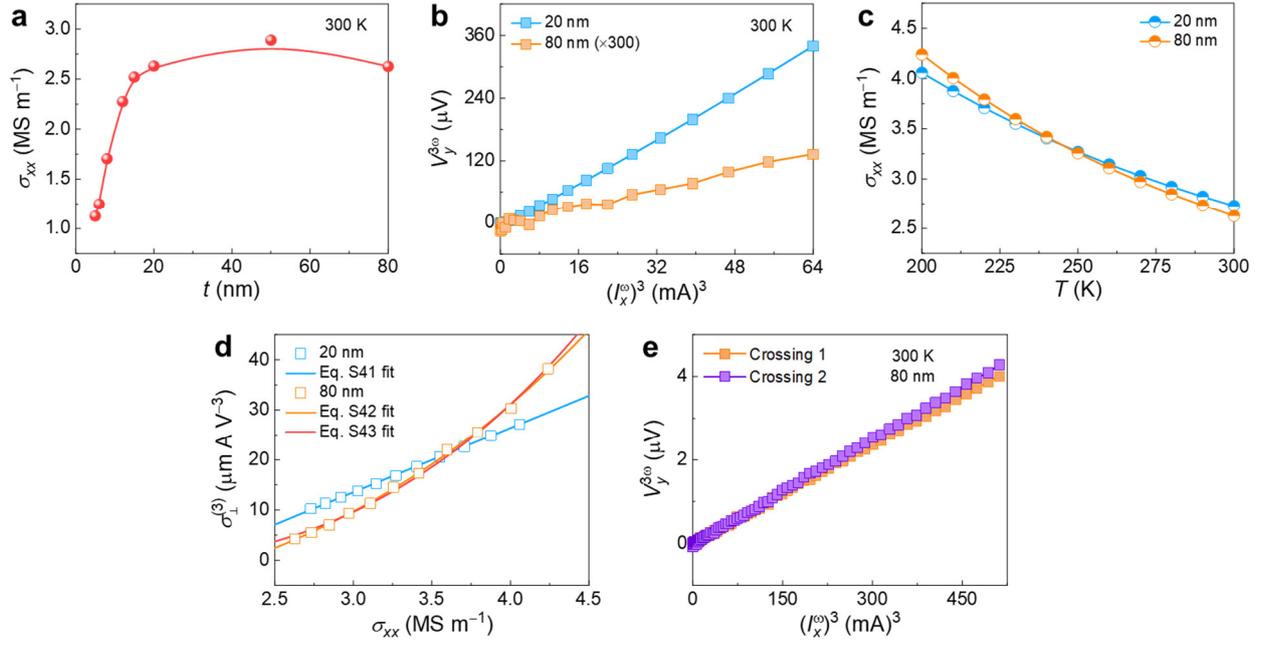

**Fig. S10 | Third-order electrical transport in thick (101)-oriented RuO$_2$ films. a**, Thickness $t$-dependent room-temperature $\sigma_{xx}$. **b**, $V_y^{3\omega}$ as a function of $(I_x^\omega)^3$ in a 20- and an 80-nm-thick film at 300 K. **c**, $T$-dependent $\sigma_{xx}$ of the two samples at 200–300 K. **d**, Relation between $\sigma_\perp^{(3)}$ and $\sigma_{xx}$ at 200–300 K for the two samples. **e**, Third-order transverse transport measured with two different Hall crossings in the same 80-nm-thick device at room temperature. $I_x^\omega$ is applied along $[\bar{1}11]^*$ in the measurement.

order anomalous Hall effect in time-reversal symmetric systems. *Phys. Rev. B* **111** (2025).